\documentclass[%
 aip,
 amsmath,amssymb,
 reprint,%
]{revtex4-2}

\usepackage{graphicx}
 \graphicspath{{figures/}}
\draft 
\usepackage{amsmath}
\usepackage{siunitx}
\usepackage[makeroom]{cancel}

\begin{document}


\title{Modeling the dominance of the gradient drift or Kelvin-Helmholtz instability
	in sheared ionospheric $\mathbf{E}\times\mathbf{B}$ flows}      



\author{C. Rathod}
\email[]{chiragr@vt.edu}
\affiliation{Kevin T. Crofton Department of Aerospace and Ocean Engineering, Virginia Tech, Blacksburg, VA 24060, USA}

\author{B. Srinivasan}
\email[]{srinbhu@vt.edu}
\affiliation{Kevin T. Crofton Department of Aerospace and Ocean Engineering, Virginia Tech, Blacksburg, VA 24060, USA}

\author{W. Scales}
\email[]{wscales@vt.edu}
\affiliation{Bradley Department of Electrical and Computer Engineering, Virginia Tech, Blacksburg, VA 24060, USA}


\date{\today}

\begin{abstract}
	Studies have shown that in sheared $\mathbf{E}\times\mathbf{B}$ flows in an inhomogeneous ionospheric plasma, the gradient drift (GDI) or the Kelvin-Helmholtz (KHI) instability may grow. This work examines the conditions that cause one of these instabilities to dominate over the other using a novel model to study localized ionospheric instabilities. The effect of collisions with neutral particles plays an important role in the instability development. It is found that the KHI is dominant in low collisionality regimes, the GDI is dominant in high collisionality regimes, and there exists an intermediate region in which both instabilities exist in tandem. For low collisionality cases in which the velocity shear is sufficiently far from the density gradient, the GDI is found to grow as a secondary instability extending from the KHI vortices. The inclusion of a neutral wind driven electric field in the direction of the velocity shear does not impact the dominance of either instability. Using data from empirical ionospheric models, two altitude limits are found. For altitudes above the higher limit, the KHI is dominant. For altitudes below the lower limit, the GDI is dominant. In the intermediate region, both instabilities grow together. Increasing the velocity shear causes both limits to be lower in altitude. This implies that for ionospheric phenomena whose density and velocity gradients span large altitude ranges, such as subauroral polarization streams, the instabilities observed by space-based and ground-based observation instruments could be significantly different. 
\end{abstract}


\maketitle 


\section{Introduction}  
\label{s:intro}

The gradient drift instability (GDI),
sometimes called the $\mathbf{E}\times\mathbf{B}$ or cross-field instability,
is a plasma interchange instability that was
initially studied in laboratory discharge experiments\cite{simon1963instability,hoh1963instability}.
A space charge or net current can develop in a perturbed,
electrostatic, inhomogeneous, weakly ionized plasma in the 
presence of background electric and magnetic fields if
there exists a difference in the effects of ion-neutral and 
electron-neutral collisions.
This results in perturbed
electric fields that drive perturbed $\mathbf{E}\times\mathbf{B}$
drifts causing instability growth. 
The GDI is a prominent 
structuring mechanism in the ionosphere\cite{fejer1980ionospheric,keskinen1983theories,tsunoda1988high}
shown to cause ionospheric
density irregularities in barium cloud experiments\cite{linson1970formation,haerendel1968artificial,mcdonald1980computer},
polar cap patches\cite{weber1984f,moen2012first,lamarche2017radar},
and diffuse aurora\cite{greenwald1974diffuse,tsunoda1982evidence,sahr1996auroral}.

The Kelvin-Helmholtz instability (KHI) is a
hydrodynamic instability that occurs at a velocity shear
interface and induces the growth of vortical motion\cite{chandrasekhar1961hydrodynamic}.
While the KHI exists ubiquitously in the universe, its primary role in the ionosphere
is to generate high altitude turbulence\cite{kintner1985status}. 
The KHI can also occur as a secondary 
instability to the GDI\cite{keskinen1988nonlinear,gondarenko1999gradient}.
In large density gradient scale length regimes, the electrostatic KHI is damped by
collisions\cite{keskinen1988nonlinear}.

The ionosphere can be categorized into three general regions (in order
of increasing altitude): the \textit{D}, \textit{E},
and \textit{F} regions. The maximum plasma density exists in the \textit{F} region. 
Ionospheric plasma throughout all of the regions is a low-$\beta$ plasma ($\approx10^{-6}$).
The ionosphere coexists with the neutral thermosphere which depletes with altitude. Therefore,
collision frequency with neutrals decreases with altitude. Thus, the effect of collisions is stronger in the 
\textit{D} and \textit{E} regions and weaker in the \textit{F} region. This work focuses on the instability development in 
the \textit{F} region where the collisions are weak enough to allow for both the GDI 
and KHI to grow.

Nonlocal linear theory on the GDI has shown that velocity shear acts
to preferentially stabilize shorter wavelength modes\cite{perkins1975velocity,huba1983shearF,huba1983shearE}.
Simulations have shown that velocity shear also prevents the extension of the 
GDI through regions of velocity shear\cite{rathod2020investigation}.
However, in simulations with similar density and velocity profiles, the KHI
has been shown to grow\cite{keskinen1988nonlinear}. This paper examines
which instability dominates under different parameter regimes.

The GDI and KHI are both important ionospheric structuring and turbulence
generation mechanisms. Ionospheric turbulence can negatively impact
radio communication signals. Therefore, understanding ionospheric turbulence
can lead to predictive space weather forecasting and improved
communication signals. Present simulation models, such as TIE-GCM\cite{dang2020development}, typically operate on global
scales with resolutions of $0.625^\circ$. The novel model presented in this paper can examine instability development
on scales of hundreds of meters.

Subauroral polarization streams (SAPS) are a specific ionospheric phenomenon
in which turbulence has been observed.  
They are characterized by a latitudinally varying 
westward $\mathbf{E}\times\mathbf{B}$ driven flow\cite{foster2002saps}
and are approximately co-located with a 
latitudinally varying density trough\cite{spiro1978ion}.
Additionally, many density and electric field
irregularities have been observed in SAPS\cite{ledvina2002first,mishin2003electromagnetic,foster2004millstone,oksavik2006first,mishin2008irregularities}.
One of the primary mechanisms hypothesized to generate these is
the GDI based
on the density gradient morphology as well as observed turbulence spectra\cite{mishin2003electromagnetic,mishin2008irregularities},
which has been corroborated by simulation\cite{rathod2020investigation}.
However, alongside the density gradients are regions of velocity shear which may induce
the KHI. Therefore, this work considers different collisional and spatial parameter regimes to determine
which of the GDI or the KHI will dominate the instability development.

This paper is organized in the following manner. Section~\ref{s:theory} provides
a brief review of linear theory on the GDI and the KHI. It also makes predictions
on the optimal GDI growth direction. Section~\ref{s:bigModel} discusses
the mathematical and computational models used in this work, as well as 
the model initialization.
Section~\ref{s:results} evaluates the predicted optimal GDI growth directions
and examines several parameters that might determine which of
the KHI or GDI are dominant. Section~\ref{s:discussion} 
interprets how the results apply to the ionosphere.
Section~\ref{s:conclusions} provides a summary of the work.
Appendix~\ref{a:model_deriv} discusses the derivation of the model in greater
detail. Appendix~\ref{a:lin} benchmarks the model to well known
linear theory for the GDI and the KHI in slab geometries.

\section{Linear Theory}  \label{s:theory}

This section provides simplified linear
growth rates used for direction and interpretation
of the simulations. Comparisons of simulation results
with established linear theory are provided
in Appendix~\ref{a:lin}.

\subsection{GDI} \label{s:GDItheory}
The growth rate
and maximum growth direction
of the GDI are
dependent on the strength of the collisions with
neutrals. The effect of collisions is characterized by
the dimensionless parameter $R$, which is
\begin{equation}
	R = \frac{\nu_{in}}{\Omega_{ci}} - \frac{\nu_{en}}{\Omega_{ce}},
\end{equation}
where $\nu$ is the collision frequency,
$\Omega$ is the gyrofrequency,
and the subscripts $i$, $e$, and $n$ denote
the ions, electrons, and neutral particles respectively. 
In the \textit{E} region, $R \gg 1$, which implies that the 
Hall current dominates\cite{sahr1996auroral}. 
In the \textit{F} region, $ 0 <  R  \ll 1 $, which implies
that the Pedersen current dominates\cite{keskinen1982nonlinear}.
When $R$ is exactly 0, the GDI is not expected to occur since no collisions
are present.

An altitude independent analytical growth rate is presented
in Eq. 16 from Ref. \onlinecite{makarevich2014symmetry} for
an arbitrary geometry.
If the magnetic field is assumed to be purely in $\mathbf{\hat{z}}$ and
the density gradient, wavenumber, and electric field are in the $xy$ plane,
that equation simplifies to 
\begin{equation}
	\gamma_{GDI} = \Big| \frac{ E }{ B L_N^g } \Big| f,
	\label{eq:gamma}
\end{equation}
where $E$ is the magnitude of the electric field, $B$ is the
magnitude of the magnetic field, $L_N^g$ is the magnitude
of the density gradient
scale length defined as $L_N^g =n | \nabla n |^{-1}$,
and $f$ is a geometric factor that is a function  
of the directions
of the density gradient, electric field, magnetic field, and wavenumber.
The electric field unit vector is defined to be
$\mathbf{\hat{e}} = \cos \theta_e \mathbf{\hat{x}}
+ \sin \theta_e \mathbf{\hat{y}}$. The wavenumber
unit vector is defined to be 
$\mathbf{\hat{k}} = \cos \theta_k \mathbf{\hat{x}}
+ \sin \theta_k \mathbf{\hat{y}}$. 
The density gradient unit vector is defined
to be 
$\mathbf{\hat{g}} = \cos \theta_g \mathbf{\hat{x}}
+ \sin \theta_g \mathbf{\hat{y}}$. 
The angles $\theta_e$, $\theta_g$, and
$\theta_k$ are in the $xy$ plane and defined to be positive
counterclockwise of $\mathbf{\hat{x}}$. Based on these definitions,
the geometric factor, $f$, is
\begin{equation}
	f =  \sin \big(\theta_k  - \theta_g \big)
	\Big[ \cos \big( \theta_e - \theta_k \big) - R \sin \big( \theta_e - \theta_k \big) \Big].
	\label{eq:f}
\end{equation}

If a random noise perturbation is
assumed to seed the instability growth, 
then the primary structures that develop will be in the fastest growing 
direction.
This direction can be found by maximizing
Eq.~\ref{eq:f} for $\theta_k$ resulting in
\begin{equation}
	\theta_{k_{\max}} = \frac{1}{2} \Big( \theta_e + \theta_g - \tan^{-1} \frac{1}{R} \Big) + l\frac{\pi}{2}
	\qquad l = \pm 1, \pm 3, \pm 5 \dots, 
	\label{eq:thetak}
\end{equation}
where $l$ is an integer denoting the periodicity of the maxima.
Note that $l$ is required to be odd based 
on the second derivative test to obtain only the maxima.
Similar representations of Eq.~\ref{eq:thetak} are
found in the literature \cite{keskinen1982nonlinear,makarevich2014symmetry}.  

Eq.~\ref{eq:gamma} is the GDI growth rate for a sufficiently collisional plasma. 
For a plasma in the inertial regime with a small $R$
such that $\nu_{in} \ll 4  |E / B L_N^g |$, 
the GDI growth rate is\cite{ossakow1978high}
\begin{equation}
	\gamma_{GDI}=  f \sqrt{\nu_{in} \frac{ E }{B L_N^g} }.
	\label{eq:GDI_inertial}
\end{equation}
Eq. 32 from Ref.~\onlinecite{makarevich2019towarda}
provides a generalized GDI growth rate with the inclusion of
inertial terms under the assumption that $|\gamma| \ll \nu_{in}$. 
In this paper, the parameter $R$ is swept through different values such
that each of these three GDI growth rate equations, (Eqs.~\ref{eq:gamma} and
\ref{eq:GDI_inertial} in this paper and Eq. 32 from Ref.~\onlinecite{makarevich2019towarda})
will be valid for different cases. There will also be cases such that none of these 
are valid due to the assumptions for the collisions.
The optimal growth direction determined by Eq.~\ref{eq:thetak}
is independent of whether the GDI is in the collisional or inertial regime.

\subsection{KHI} \label{s:KHItheory}

For a fluid with a density and velocity interface, 
the KHI growth rate is
\begin{equation}
	\gamma_{KHI} = \frac{ \rho_1 \rho_2 k ( V_1 - V_2 ) }{ (\rho_1 + \rho_2)^2},
	\label{eq:KHI}
\end{equation}
where $\rho$ is the mass density, $V$ is the velocity, $k$ is the wavenumber,
and the subscripts denote values on either side of the interface\cite{chandrasekhar1961hydrodynamic}.
Eq.~\ref{eq:KHI} applies to a sharp interface modeled by a Heaviside function. For a general hydrodynamic shear flow
with a gradually transitioning interface, 
the growth rate decreases with a larger velocity gradient scale length
or with a smaller density gradient scale length\cite{wang2010combined}.

The effect of collisions with neutral particles plays an important
role for the KHI in the electrostatic regime.
Contrary to the hydrodynamic description
of the KHI\cite{wang2010combined}, decreasing the density gradient scale length can lead to a regime in which
the KHI growth rate increases with higher collisionality\cite{hysell2004collisional}. 
The KHI in this regime is sometimes called the collisional shear instability. 
For a regime with large density gradient scale lengths, increased collisionality
has a stabilizing effect on the KHI growth rate\cite{keskinen1988nonlinear}.
This latter regime is what is considered in this work primarily
due to its applicability to subauroral polarization streams in the \textit{F} region ionosphere.

Additionally, the magnetic field in this work is perpendicular to the velocity shear direction
and therefore does not impact the instability growth\cite{chandrasekhar1961hydrodynamic}.

\section{Model} \label{s:bigModel}

\subsection{Mathematical Model} \label{s:model}

A novel mathematical model focused on studying the temperature
gradient instability\cite{hudson1976temperature} theoretically
is presented by Ref.~\onlinecite{keskinen2004midlatitude}. 
The work presented here adapts the concepts from that model
to develop a numerical framework to study the KHI and the GDI.
Production and loss terms as well as the temperature source terms
are not considered as they have a negligible impact on the development
of the instabilities being modeled. This work, in contrast with
Ref.~\onlinecite{keskinen2004midlatitude}, includes inertial terms
to better understand the KHI, the inertial GDI, and the
nonlinear turbulence cascade.

The \textit{F} region plasma is modeled electrostatically 
as a 2D fluid. Only
motion perpendicular to the magnetic field is considered. The continuity,
momentum, and energy equations, respectively, for an arbitrary
species $\alpha$, are
\begin{gather}
	\frac{ \partial n_\alpha }{ \partial t } + 
	\nabla \cdot ( n_\alpha \mathbf{V}_\alpha ) 
	= D \nabla^2 n_\alpha \label{eq:cont} \\
	n_\alpha \Big( \frac{\partial}{\partial t} + \mathbf{V}_\alpha 
	\cdot \nabla \Big) \mathbf{V}_\alpha = 
	\frac{q_\alpha n_\alpha}{m_\alpha} ( 
	\mathbf{E} + \mathbf{V}_\alpha \times \mathbf{B}) \nonumber \\
	- \frac{\nabla P_\alpha}{m_\alpha} - 
	n_\alpha \nu_{\alpha n} ( \mathbf{V}_\alpha
	- \mathbf{u}) \label{eq:mtm} \\
	\frac{3}{2} n_\alpha k_B 
	\frac{\partial T_\alpha}{\partial t} 
	+ \frac{3}{2} n_\alpha k_B 
	\mathbf{V}_\alpha \cdot \nabla T_\alpha 
	+ n_\alpha k_B T_\alpha 
	\nabla \cdot \mathbf{V}_\alpha = 0 \label{eq:temp}
\end{gather}
where $D$ is an numerical diffusion constant, $\mathbf{u}$ 
is the neutral wind speed, and $\nu_{\alpha n}$ is the collision frequency 
between species $\alpha$ and neutral particles.
The only species evolved in this model are ions and electrons;
the effect of the neutral particles is considered in the collisional momentum source term. 
The equation of state is the ideal gas law. The collision frequencies are 
calculated using $\nu_{\alpha n} = n_n V_{Th_\alpha} \sigma_{\alpha n}$,
where $n_n$ is the neutral number density, $V_{Th_\alpha}$ is the thermal velocity,
and $\sigma_{\alpha n}$ is the collision cross-section.
The energy equation, Eq.~\ref{eq:temp}, is included in the model and
is solved computationally, but has a negligible impact on these instabilities
in this parameter regime. Therefore, the emphasis is placed on Eqs.~\ref{eq:cont}
and~\ref{eq:mtm}.

The velocities are found from Eq.~\ref{eq:mtm} and split into
zeroth and first order in terms of the form $\nu_{\alpha n}/\Omega_{c\alpha}$ or
$(\partial/\partial t + \mathbf{V} \cdot \nabla)/\Omega_{c\alpha}$,
where $\Omega_{c\alpha}$ is the gyrofrequency. The resulting
velocities are
\begin{align}		
	\mathbf{V}_e^0 &= \frac{\mathbf{E}\times\mathbf{B}}{B^2} 
	+ \frac{ \nabla P_e \times \mathbf{B} }{ e B^2 n_e}
	\label{eq:ve0}\\
	\mathbf{V}_e^1 &= -\frac{\nu_{en}}{\Omega_{ce}} 
	\frac{\nabla P_e}{e n_e B} -
	\frac{\nu_{en}}{B \Omega_{ce}} 
	\Big( \mathbf{E} + \mathbf{u} \times \mathbf{B} \Big) \nonumber \\
	&\ - \frac{1}{B\Omega_{ce}} 
	\Big( \frac{\partial}{\partial t} + 
	\mathbf{V}_{\mathbf{E}\times\mathbf{B}} \cdot \nabla \Big)
	\mathbf{E} \label{eq:ve1}\\
	\mathbf{V}_i^0 &= \frac{\mathbf{E}\times\mathbf{B}}{B^2} 
	- \frac{ \nabla P_i \times \mathbf{B} }{ e B^2 n_i}
	\label{eq:vi0}\\
	\mathbf{V}_i^1 &=  -\frac{\nu_{in}}{\Omega_{ci}} 
	\frac{\nabla P_i}{e n_i B} +
	\frac{\nu_{in}}{B \Omega_{ci}} 
	\Big( \mathbf{E} + \mathbf{u} \times \mathbf{B} \Big) \nonumber \\
	&\ + \frac{1}{B\Omega_{ci}} 
	\Big( \frac{\partial}{\partial t} + 
	\mathbf{V}_{\mathbf{E}\times\mathbf{B}} \cdot \nabla \Big)
	\mathbf{E}, \label{eq:vi1}
\end{align}
where the zeroth order velocities are a combination of the $\mathbf{E}\times\mathbf{B}$
and diamagnetic drifts. 

The zeroth order velocities are used in the continuity and energy equations.
Because the plasma is electrostatic, the electric field is defined 
as $\mathbf{E}=-\nabla \phi$. 
Due to the constant
magnetic field, the diamagnetic drift term does not impact the continuity equation
and therefore only the $\mathbf{E}\times\mathbf{B}$ drift advects the plasma. 
The plasma is assumed to be quasineutral such that $n_i = n_e = n$, and thus, only one
continuity equation is needed.
Furthermore, the current is closed using $\nabla \cdot \mathbf{J} = 0$ where
$\mathbf{J} = n e (\mathbf{V}_i - \mathbf{V}_e )$ where both the zeroth and
first order velocities are used to calculate $\mathbf{J}$. 
Therefore, the relevant equations, continuity and current closure, are 
\begin{equation}
	\frac{ \partial n }{ \partial t } 
	- \frac{\nabla \phi \times \mathbf{B}}{B^2}
	\cdot \nabla n
	= D \nabla^2 n \label{eq:contFinal} 
\end{equation}
\begin{multline}
	\nabla \cdot \Big( n \nabla 
	\frac{\partial \phi}{\partial t}  \Big) 
	=  \Big( \frac{1}{\Omega_{ci}} 
	+ \frac{1}{\Omega_{ce}} \Big)^{-1} 
	\bigg[ - \frac{\nu_{in}}{e\Omega_{ci}} \nabla^2 P_i 
	+ \frac{\nu_{en}}{e\Omega_{ce}} \nabla^2 P_e \\
	+ \Big( \frac{\nu_{in}}{\Omega_{ci}}
	+ \frac{\nu_{en}}{\Omega_{ce}}\Big) 
	\Big( \mathbf{u}\times\mathbf{B} \cdot \nabla n  
	-\nabla \cdot [ n \nabla \phi] \Big)\bigg] \\
	- \nabla \cdot \Big( n \mathbf{V}_{\mathbf{E}\times\mathbf{B}} \cdot \nabla \nabla \phi\Big).
	\label{eq:divJ0}
\end{multline}
The  $\mathbf{u}\times\mathbf{B}$ term can be considered an
applied electric field, $\mathbf{E}_0$. 
This field is the result of a Lorentz transformation into the frame
of reference of the neutral particles. Note that the numerical diffusion 
term in Eq.~\ref{eq:contFinal} provides isotropic diffusion to mitigate numerical error,
preferentially acting on higher wavenumber modes.

Traditionally, fundamental physics simulations of plasma instabilities
are required to start in an equilibrium state. This is done by 
initializing all 
but one of the variables and allowing the remaining variable to be solved for
such that the system is in an equilibrium. However, the governing equations 
and collision frequencies are highly coupled.
Assigning any one variable as the equilibrium variable
would have significant consequences for the growth of the GDI or the KHI.
Therefore, to maintain an equilibrium,
a perturbed model is developed such that the background does not change
and only the perturbation is evolved. This inherently makes
the assumption that the time scales of turbulence evolution are much faster
than the time scales of the background dynamics. This can also be seen as a method of accounting 
for other physics, such as motion parallel to the magnetic field, that exist in nature to maintain 
a quasi-equilibrium. The variables $\phi$, $n$, $T_i$, and $T_e$ are split into
background and perturbed quantities, e.g., $\phi = \phi_{bg} + \phi_p$. When substituting
these terms into Eqs.~\ref{eq:contFinal} and~\ref{eq:divJ0}, terms containing 
only background quantities are removed. Thus, only terms with perturbed
quantities remain. Based on the background geometry of the problem, as
discussed in Section~\ref{s:ICs}, the
continuity and energy residuals already satisfy the criterion of 
removing the background quantities. Thus, this method
only needs to be applied to the continuity temporal derivative,
the continuity source term, and the entire current closure equation. The 
resulting perturbed set of equations are
\begin{equation}  
	\frac{\partial n_p}{\partial t} - 
	\frac{\nabla \phi \times \mathbf{B}}{B^2}
	\cdot \nabla n
	= D \nabla^2 n_p
	\label{eq:contPert}
\end{equation}  
\begin{widetext}
\begin{multline} 
	\nabla \cdot \Big( n \nabla 
	\frac{\partial \phi_p}{\partial t} \Big) =
	\Big( \frac{1}{\Omega_{ci}} 
	+ \frac{1}{\Omega_{ce}} \Big)^{-1} 
	\bigg[ - \frac{\nu_{in}}{e\Omega_{ci}} \nabla^2 P_{i_p}  
	+ \frac{\nu_{en}}{e\Omega_{ce}} \nabla^2 P_{e_p} 
	+ \Big( \frac{\nu_{in}}{\Omega_{ci}}
	+ \frac{\nu_{en}}{\Omega_{ce}}\Big) 
	\Big( \mathbf{u}\times\mathbf{B} \cdot \nabla n_p   
	-\nabla \cdot [ n_p \nabla \phi + n_{bg} \nabla \phi_p] \Big)\bigg] \\
	- \nabla \cdot \Big( n_{bg} \mathbf{V}_{\mathbf{E}\times\mathbf{B}_{bg}} \cdot \nabla \nabla \phi_p + n_{bg} \mathbf{V}_{\mathbf{E}\times	
	\mathbf{B}_p}\cdot \nabla \nabla \phi + n_p \mathbf{V}_{\mathbf{E}\times\mathbf{B}} \cdot \nabla \nabla \phi \Big).
	\label{eq:divJpert}
	\end{multline}
\end{widetext}

The derivation of the model is discussed
in greater detail in Appendix~\ref{a:model_deriv}.

Spatially, the equations are discretized using a pseudo-spectral method
with a Fourier basis \cite{canuto2012spectral}. This converts the spatial
derivatives into algebraic expressions, resulting in a set of temporal
ordinary differential equations (ODEs). Periodic boundary conditions in all 
dimensions are required for the basis chosen for this discretization method. 
The temporal derivative term in Eq.~\ref{eq:divJpert}, 
$\nabla \cdot ( n \nabla  \partial \phi_p / \partial t)$, 
cannot be isolated. An iterative method is used to 
calculate $\partial \phi_p / \partial t$. This results in a linear system
of temporal ODEs that are integrated using a four stage
fourth order Runge-Kutta method\cite{hirsch1990numerical}. 

\subsection{Background and Initial Conditions}\label{s:ICs}   

For all of the simulations presented in this paper, the
functional forms of the background variables are the same.

The background density is defined by a set
of two hyperbolic tangent functions of the form
\begin{equation}
	n_{bg} (x) =n_0\bigg( c + \sum_{i=1}^{2}
	a_i \tanh \Big[ \frac{ x - x_{N_i} }{L_{N}} \Big] \bigg),
	\label{eq:n}
\end{equation}
where $n_0$ is a reference density,
$L_N$ is a length scaling factor, $x_N$ defines the location of the density
interfaces, and the constants $a$ and $c$ determine the levels
of each region of the hyperbolic tangents. All of the simulations
in Section~\ref{s:results} use
$n_0=10^{11}$~\si{m^{-3}}, $a_1=-0.75$, $a_2=0.75$ and $c=1$. The parameters 
$L_N$ and $x_{N_i}$ vary.

Similarly, the background velocity is defined by four 
hyperbolic tangent functions of the form
\begin{equation}
	V_{y_{bg}} (x) = \frac{V_0}{2} \sum_{i=1}^{4}
	b_i \tanh \Big( \frac{ x - x_{V_i} }{ L_V } \Big),
	\label{eq:vTanh}
\end{equation}
where $V_0$ is a reference velocity, $L_V$ is a length
scaling factor, and $x_V$ is the location
of the velocity interfaces.
Because the background velocity is purely $\mathbf{E}\times\mathbf{B}$ 
driven, the two additional hyperbolic tangent functions are included
to ensure periodicity in the electric potential, which is calculated
using $\phi_{bg}(x) = B \int V_{y_{bg}} (x) dx$.
The constants $b_i$ are defined to be $b_1=1$, $b_2=-1$,
$b_3=-1$, and $b_4=1$. To maintain periodicity, $x_{V_{3,4}}$
are defined as
$x_{V_3} = L_x - x_{V_2}$ and $x_{V_4} = L_x - x_{N_1}$,
where $L_x$ is the domain length in $x$. All of the simulations use 
$L_V=10$ km and $x_{V_2}=700$ km. The parameters $V_0$ and $x_{V_1}$ vary.

The background ion and electron temperatures are constants with
$T_{i_{bg}} = T_{e_{bg}} = 1000$ K. 
The neutral species is atomic oxygen (O) and the ion
species is singly ionized atomic oxygen (O$^+$). The background 
magnetic field is $B=\SI{5e-5}{T}$. The density is initialized 
with a random noise perturbation of  $10^{-6} n_0$.
The domain length, number of grid points, numerical diffusion constant,
neutral number density,
and neutral wind vary between simulations.
All of the parameters have been chosen to reasonably align 
with observed values in subauroral polarization streams or
the \textit{F} region ionosphere in general. The chosen parameters
show that the fluid approximation is valid
with the 
Debye length (\SI{4e-3}{m}), the ion (\SI{3}{m}) and 
electron (\SI{1e-2}{m}) gyroradii,
the skin depth (\SI{10}{m}), and ion inertial length (\SI{1.8e3}{m})
not being adequately resolved by the smallest used grid size of \SI{732}{m}.
Since the phenomena considered are at much larger scales, the inherent model
assumptions cannot capture kinetic physics.

\section{Results} \label{s:results}

\subsection{Optimal GDI Growth Direction as a Diagnostic}

A set of simulations is run to test 
the validity of Eq.~\ref{eq:thetak} in predicting
the direction of GDI growth. This is a useful diagnostic
for Sections~\ref{s:collisions} and \ref{s:vsLoc} to 
verify that some of the instability structures that develop are
the GDI based on the direction of growth. 
It is tested by changing the direction of the applied electric field.
The electric field for each case is initialized
such that the magnitude remains constant but the angle is varied from 0 to $7\pi/4$
in increments of $\pi/4$ for a total of 8 cases. 
Since the applied electric field
is based on the neutral wind, the neutral wind is initialized as
$\mathbf{u} = u \cos ( \theta_e + \pi/2 )\mathbf{\hat{x}} + u \sin (\theta_e + \pi / 2) \mathbf{\hat{y}}$,
where $u=500$~m/s. Since the initial density perturbation is 
random white noise, the dominant GDI that grows will have the highest growth
rate, as per Eq.~\ref{eq:gamma}. This highest growth rate should occur for a wavenumber
with an angle defined by Eq.~\ref{eq:thetak}.

The domain length is
$L = 500 \times  250$ km with $256\times 512$ grid points.
The parameters for the background density profile are $L_N = 30$ km,
$x_{N_1}=125$ km, and $x_{N_2}=375$ km. The background velocity profile is turned
off by setting $V_0=0$ m/s. The numerical diffusion constant
is $10^3$ \si{m^2/s}. The neutral number density is $n_n=10^{14}$ \si{m^{-3}}.

\begin{figure}[!htb]  
	\includegraphics[width=\linewidth]{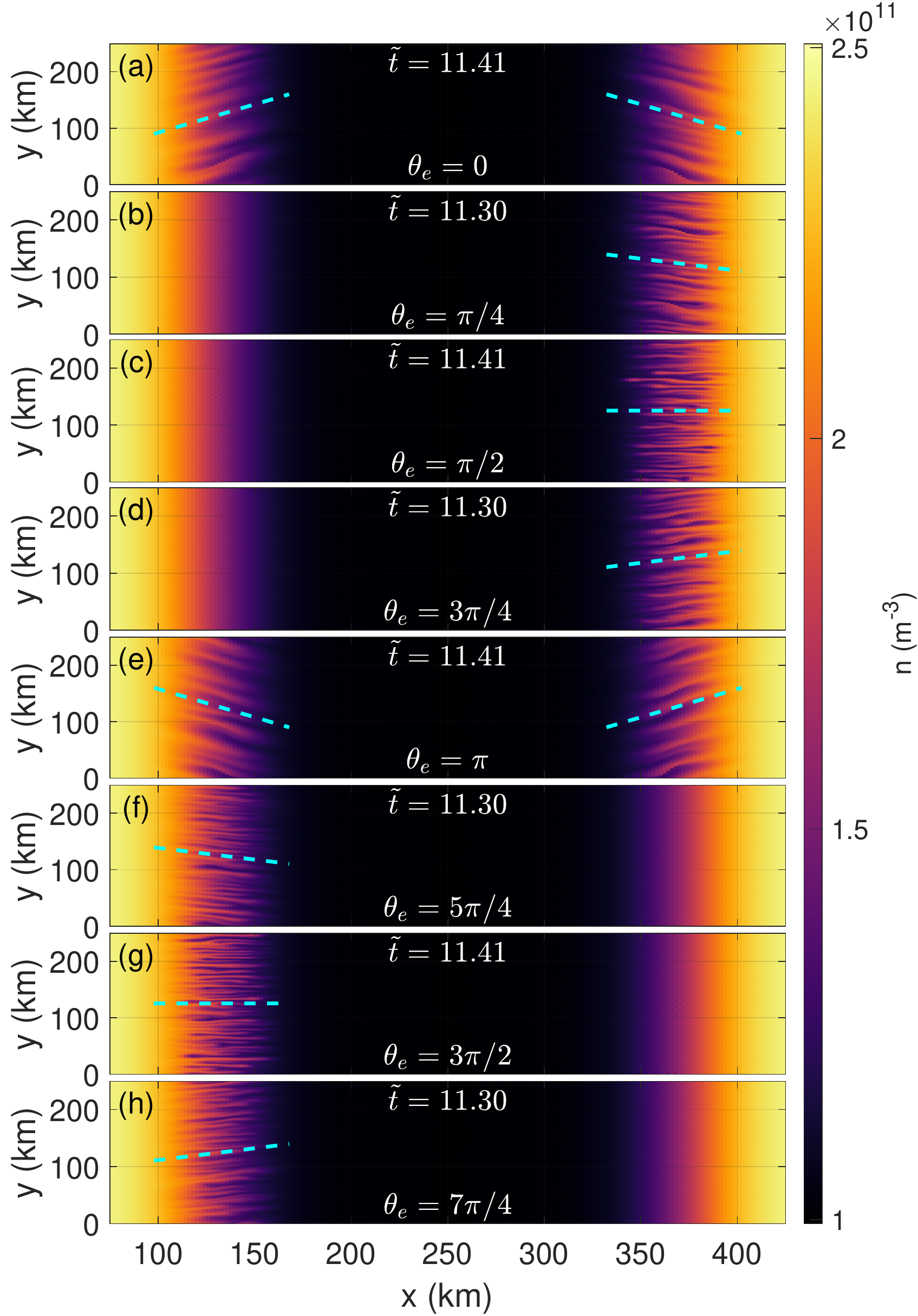}
	\caption{Density plots showing the development of the GDI for
		$\theta_e$ from 0 to $7\pi/4$. The dashed cyan lines 
		indicate the direction of predicted visual growth based
		on the Eq.~\ref{eq:thetak}. Note that the visual growth of the GDI
		is perpendicular to the optimal wavenumber direction, $\theta_{k_{\max}}$.
		For all cases, the GDI extends in the predicted directions. 
		Depending on the value of $\theta_e$, the growth at a particular interface,
		based on Eq.~\ref{eq:f}, may be significantly stronger than at the other 
		interface. For this reason, the GDI only grows at
		the right density interface for Panels (b-d) and at the left 
		density interface for Panels (f-h).
			\label{f:GDIkDir}
		}
\end{figure}

Figure~\ref{f:GDIkDir} shows the results of this 
set of simulations. The times
are shown non-dimensionally with $\tilde{t}=u f_{\max} t /L_N^g$,
where $L_N^g$ is the minimum density gradient scale length
and $f_{\max}$ is the larger geometric factor (from Eq.~\ref{eq:f})
between the left and right density interfaces. 
In all of the simulations,
the GDI is shown to grow as ``fingers" extending in
both directions from the location of 
minimum density gradient scale length. The dashed cyan lines 
on each plot indicate the direction of predicted GDI growth.
Note that the direction of the visual GDI growth, i.e., the ``fingers", is perpendicular to
the GDI propagation direction. Therefore, the slopes of the dashed cyan lines 
represent the angles of $\theta_{k_{\max}} \pm \pi/2$. Because there are two density interfaces,
the effect of changing $\theta_g$ is also considered. The left density interfaces
have $\theta_g = \pi$ and the right density interfaces have $\theta_g=0$. 
Based on $\theta_g$ and $\theta_e$, the GDI visually extends in the 
directions predicted by the dashed cyan lines validating the linear theory
prediction of Eq.~\ref{eq:thetak} (note that further benchmarking of
the GDI to linear theory is provided in Appendix~\ref{a:lin}). 
The geometric factor, $f$, 
is different for each density interface 
because of the different directions of $\theta_g$. 
Because of this, one interface may have a larger growth rate
than the other and therefore dominates the growth.
This is seen in Figures~\ref{f:GDIkDir}(b-d) where the GDI
grows on the right interface and in Figure~\ref{f:GDIkDir}(f-h)
where the GDI grows on the left interface. 
Given a longer period of time, the GDI is expected to slowly grow
at the opposite interface.
However, Figures~\ref{f:GDIkDir}(c)
and~\ref{f:GDIkDir}(g) show special cases in which the GDI does not grow
on the opposite interface because $f=0$ at that interface.

\subsection{Effect of Collisions} 
\label{s:collisions}

Section~\ref{s:GDItheory}
describes the importance of the parameter $R$ on the GDI growth through Eqs.~\ref{eq:gamma}
and \ref{eq:f}. 
For large density gradient scale lengths, the KHI grows more quickly at low $R$\cite{keskinen1988nonlinear}.
The effect of varying $R$ has previously been studied 
on the independent development of the GDI\cite{mitchell1985simulation} 
and the KHI\cite{keskinen1988nonlinear}, but not on the combined
effect of these two instabilities.

Simulations are performed to sweep through different $R$ to understand
the effect of collisions on instability development looking at
the combination of the GDI and the KHI. Note that $R$ still needs to be sufficiently
small such that the assumptions made in Section~\ref{s:model} for an \textit{F} region plasma
are not violated. The parameter $R$ is changed by varying the neutral number density from 
\SI{5e12}{m^{-3}} to \SI{3e13}{m^{-3}} which corresponds to a range of $R$ from
\num{4.62e-6} to \num{2.95e-5}. These $R$ correspond to an altitude
range of approximately 700 km down to 530 km (from Figure~\ref{f:RvAlt}).  
These $R$ values are chosen to provide approximate limits as to when either the GDI
or the KHI dominates over the other for this velocity profile. In this paper, the word dominate refers to 
when one instability visually grows much more quickly than the other.

The domain length is $1500\times 1000$ km with $1024 \times 256 $ grid points. 
The parameters for the background density profile are
$L_N = 50$ km, $x_{N_1}=220$ km, and $x_{N_2}=1200$ km. The
parameters for the background velocity profiles are
$V_0=-1000$ m/s and $x_{V_1}=220$ km. The numerical diffusion
constant is $10^4$ \si{m^2/s}. The velocity magnitude corresponds to
observed values in subauroral polarization streams (SAPS)\cite{foster2004millstone}.
The neutral wind is set to 0 m/s.

\begin{figure}[!htb]   
	\includegraphics[width=\linewidth]{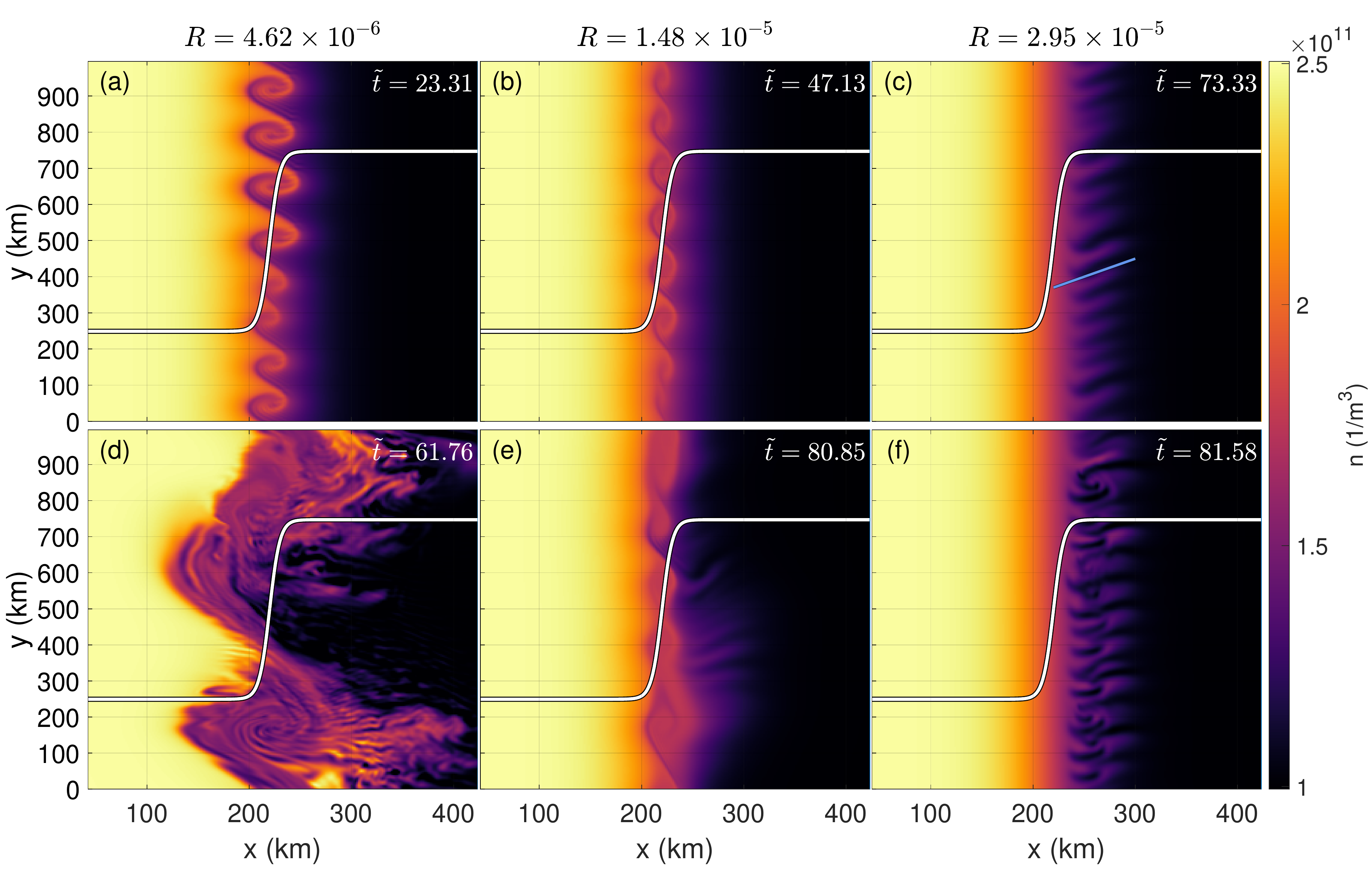}  
	\caption{Density color plots showing instability development for different
		$R$ increasing to the right. 
		Panels (a-c) display the early time growth whereas Panels (d-f) show the late time nonlinear development.
		The white
		curve represents the absolute value of the background velocity profile with $V_0=-10^3$ m/s. The transition from
		a pure KHI to a pure GDI is shown as $R$ increases. 		
		Note how the vortical structures gradually disappear and how the ``finger-like" structures
		gradually increase as $R$ increases. In Panel (c), the blue line indicates the expected
		direction of GDI growth based on Eq.~\ref{eq:thetak}.
	\label{f:GDIKHI}
	}
\end{figure}

Figure~\ref{f:GDIKHI} shows the resulting density color plots of
the instability development with increasing $R$ from left to right.
The white curve represents the absolute value of the velocity profile 
such that there is a velocity of 0 m/s on the left of the 
velocity interface and a velocity of -1000 m/s on the right of the velocity interface.
The times are shown non-dimensionally with $\tilde{t} = |t V_0 / L_N^g|$. The top panels
of Figure~\ref{f:GDIKHI} show the early time development of the instabilities and the bottom panels
show the late time nonlinear development of the instabilities.

Figures~\ref{f:GDIKHI}(a-b) show the classical vortical KHI structures at the velocity interface. 
The linear growth of Figure~\ref{f:GDIKHI}(a) reasonably matches established linear theory 
based on Figure 3 from Ref.~\onlinecite{keskinen1988nonlinear}. Further discussion of
the benchmarking of the KHI is presented in Appendix~\ref{a:lin}.
Note that Figure~\ref{f:GDIKHI}(b) shows the KHI having a significantly smaller amplitude than that of Figure~\ref{f:GDIKHI}(a)
because the KHI becomes more damped with increasing collision frequency in regimes 
of large density gradient scale lengths\cite{keskinen1988nonlinear}.
For these two cases, the KHI is the dominating instability early in time. Figure~\ref{f:GDIKHI}(c)
shows the ``finger-like" structures of the GDI growing in the direction predicted by
Eq.~\ref{eq:thetak}, which is shown by the blue line. At early time, only the case with the highest collisionality
shows any growth of the GDI.

Figure~\ref{f:GDIKHI}(d) shows that the lowest collisionality case develops
into a set of nonlinear KHI vortices with cascading turbulence but no indication of any GDI development.
Figures~\ref{f:GDIKHI}(e), however, shows that the GDI begins to develop at late
times. Figure~\ref{f:GDIKHI}(f) shows the nonlinear evolution
of the GDI. At late times, the GDI structures curve because the entire instability region
is moving in the negative $\mathbf{\hat{y}}$ direction. An additional feature that develops
is a secondary KHI, as shown by the curving ``mushroom-like" structures. For larger $R$,
the secondary KHI are expected to become increasingly damped\cite{mitchell1985simulation}.
Therefore, for low $R$, the KHI dominates, and for high $R$, the GDI dominates.
In the intermediate region, both instabilities grow in tandem.

Figure~\ref{f:GDIKHI} shows that the collisional parameter $R$ plays an
important role in determining which instability dominates
the growth. A parameter that can enhance the growth of both instabilities
is the velocity magnitude, $V_0$. A set of simulations is run with varying $R$ by
varying the neutral density from \SI{7e12}{m^{-3}} to \SI{6e13}{m^{-3}}, which
corresponds to a range of $R$ from \num{6.89e-6} to \num{5.90e-5}. 
These $R$ correspond to an altitude
range of approximately 670 km down to 490 km (from Figure~\ref{f:RvAlt}).
In these simulations,
the background velocity is set to $V_0 = -2000$ m/s.
Velocities this high can occur in subauroral ion drifts (SAID), which are 
latitudinally narrower than SAPS\cite{anderson1991ionospheric}.
The remaining
parameters are the same as those used in Figure~\ref{f:GDIKHI} except for the numerical diffusion
constant which is set to \SI{4e4}{m^2/s}.

\begin{figure}[!htb]
	\includegraphics[width=\linewidth]{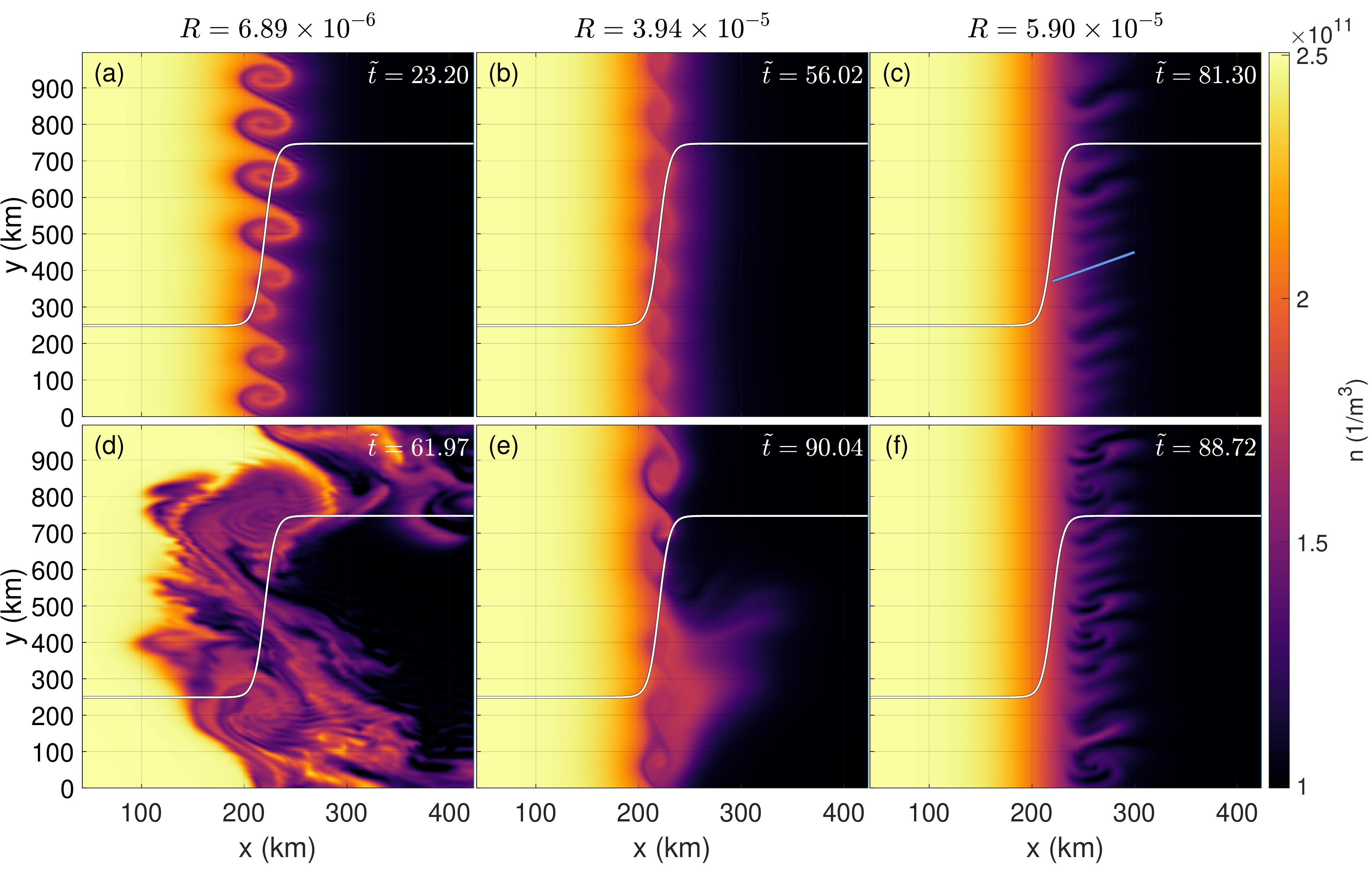}
	\caption{Density color plots showing instability development for different
		$R$ increasing to the right. 
		Panels (a-c) show the early time development and Panels (d-f) show the late time development.
		The white
		curve represents the background velocity profile with $V_0=-2000$ m/s. 		
		The qualitative behavior of the transition from the KHI to the GDI 
		matches that of Figure~\ref{f:GDIKHI}.
		Due to the higher background velocity, the transition from the KHI to the 
		GDI occurs at a higher $R$. Thus, the KHI is more
		sensitive to increases in the velocity than the GDI is. 
		Panel (c) shows the GDI dominating with the blue line representing
		the expected growth direction, which is the same as that of Figure~\ref{f:GDIKHI}(c).
	\label{f:GDIKHIV1e4}
	}
\end{figure}

Figure~\ref{f:GDIKHIV1e4} shows the resulting instability development
with $R$ increasing to the right. The time is presented non-dimensionally
with $\tilde{t}= | t V_0 / L_N^g|$. The top panels
are early in time and the bottom panels are late in time.
Qualitatively, the results are consistent with the transition from the KHI
to GDI with increasing $R$ from Figure~\ref{f:GDIKHI}.  
The impact of a higher background velocity increases the region of KHI dominance. 
Because of this, the transition between the KHI and the GDI
occurs at higher $R$. Once $R$ is sufficiently high, the GDI dominates and shows
the same general structure as in Figure~\ref{f:GDIKHI}(c).

\subsection{Effect of Velocity Shear Location}  \label{s:vsLoc}

It is clear that the dominant instability transitions from the KHI
to the GDI with increasing collisionality or decreasing velocity shear.
It is possible that the background velocity profile chosen for
Figure~\ref{f:GDIKHI} may not be ideally conducive to GDI growth.
Velocity shear has a stabilizing effect on the GDI that preferentially
impacts shorter wavelength modes \cite{perkins1975velocity,huba1983shearF,huba1983shearE}.
The region of velocity shear overlaps the region of the density gradient.
In SAPS however, the location of the velocity shear does not necessarily have to be co-located
with the density gradient\cite{rathod2020investigation}.
If the 
region of velocity shear is sufficiently far from the density gradient, then the
GDI may become more prominent even in the low collisionality regime. Note that the region of constant
velocity must overlap the density gradient to induce the GDI. Therefore, in this case,
the velocity shear location needs to be moved to the left, i.e., $x_{V_1}$ needs to be smaller.

A set of simulations is run examining different locations of the velocity shear.
Two cases are run with the parameter $x_{V_1}$ set to 200 km and 150 km. 
The remaining parameters are the same as those in Figure~\ref{f:GDIKHI}(a) except
with a resolution of $2048 \times 512$ grid points.

\begin{figure}[!htb]
	\includegraphics[width=\linewidth]{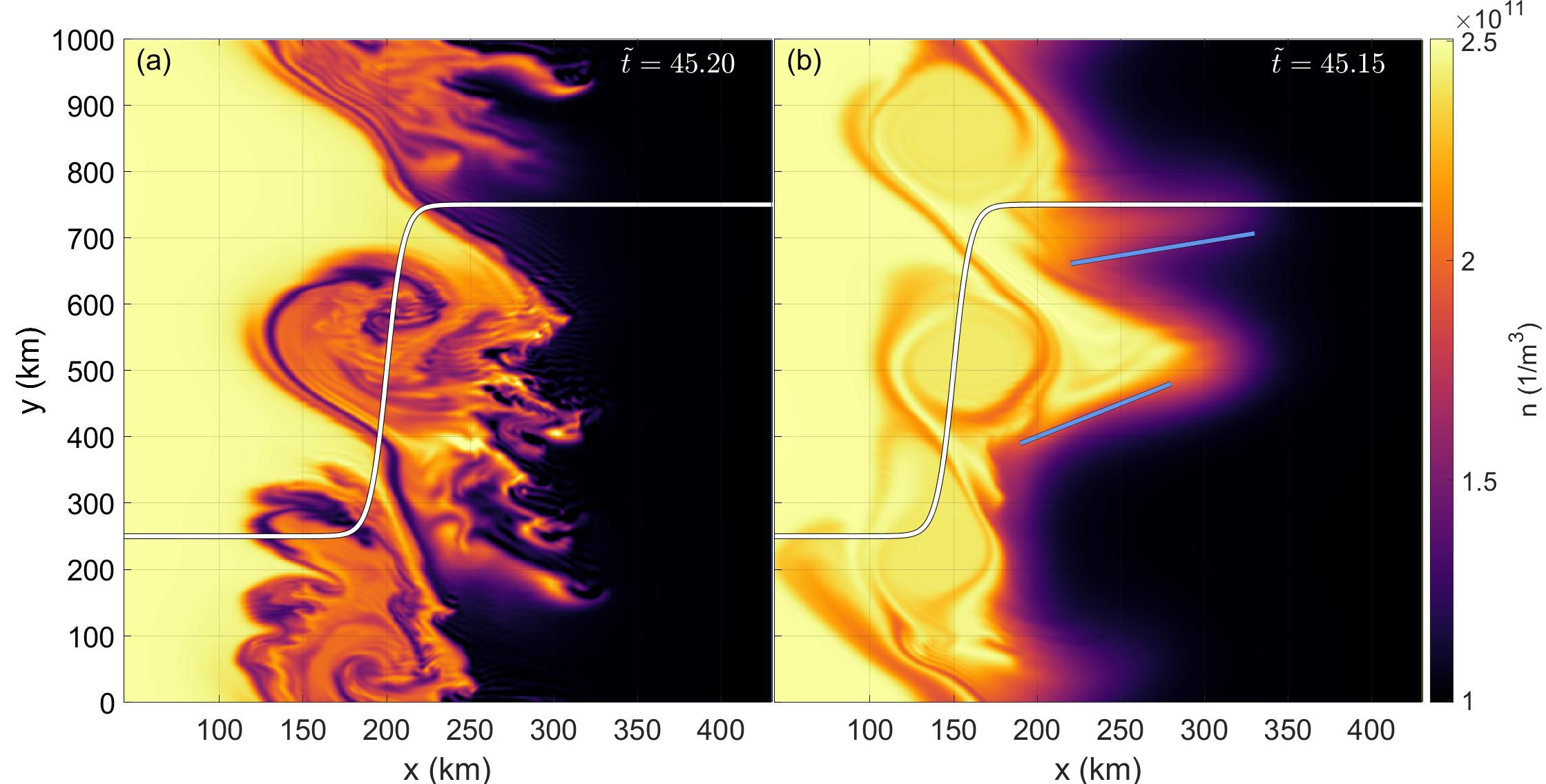}
	\caption{Density color plots showing KHI development
		for different velocity shear locations ($x_{V_1}$). 
		The white curve represents the velocity profile.
		Panel (a) shows that the KHI is the dominant instability.
		Panel (b) has a velocity profile sufficiently far away from
		the density gradient. In addition to the KHI that grows at the velocity
		shear interface, there is a long wavelength growth on the right that occurs in the direction
		of expected GDI growth (light blue line).
			\label{f:shear}
		}
\end{figure}

Figure~\ref{f:shear} shows the results of these simulations in
the late time nonlinear phase of instability growth. The time
is presented non-dimensionally as $\tilde{t} = | t V_0 / L_N^g|$. 
At early times, Figure~\ref{f:shear} looks similar to the KHI growth in
Figure~\ref{f:GDIKHI}(a).
Figure~\ref{f:shear}(a) shows a case where the velocity profile
is shifted slightly to the left, where the KHI is still the dominant instability. 
Note that the small structures extending to the bottom right from the KHI vortices
are the result of a turbulence cascade.
In Figure~\ref{f:shear}(b), the velocity shear is sufficiently far from the density
gradient to induce some sort of GDI growth. KHI vortices can be seen at the location
of velocity shear. Extending from the right of these vortices are two long wavelength structures.
Because the KHI vortex is generally circular, the value of $\theta_g$ can vary from
$\pi/2$ to $\pi$. The blue lines indicate the directions of predicted
GDI growth; the top line is uses $\theta_g=3\pi/4$ and the bottom line uses
$\theta_g=\pi$. Note that $\theta_e=0$ for both lines.
Very quickly afterwards, the GDI enters the nonlinear regime
and becomes difficult to differentiate from the KHI.
Therefore, the GDI appears to grow in situations where the velocity shear location
is sufficiently far from the density gradient even in low collisionality regimes.

Based on Eq.~\ref{eq:gamma}, adding an applied homogeneous electric field through the neutral wind can
increase the growth rate of the GDI. Based on the geometry of Figure~\ref{f:shear},
the electric field direction that would maximize the growth rate is in the $-\mathbf{\hat{y}}$ 
direction, which corresponds to a neutral wind in the $\mathbf{\hat{x}}$ direction.
A set of simulations are run with the same parameters as those in Figure~\ref{f:shear}
but with an $\mathbf{\hat{x}}$ direction neutral wind of 500 m/s. This magnitude is chosen
because it represents the upper range of what would exist in the ionosphere \cite{buonsanto1999ionospheric}.

\begin{figure}[!htb] 
	\includegraphics[width=\linewidth]{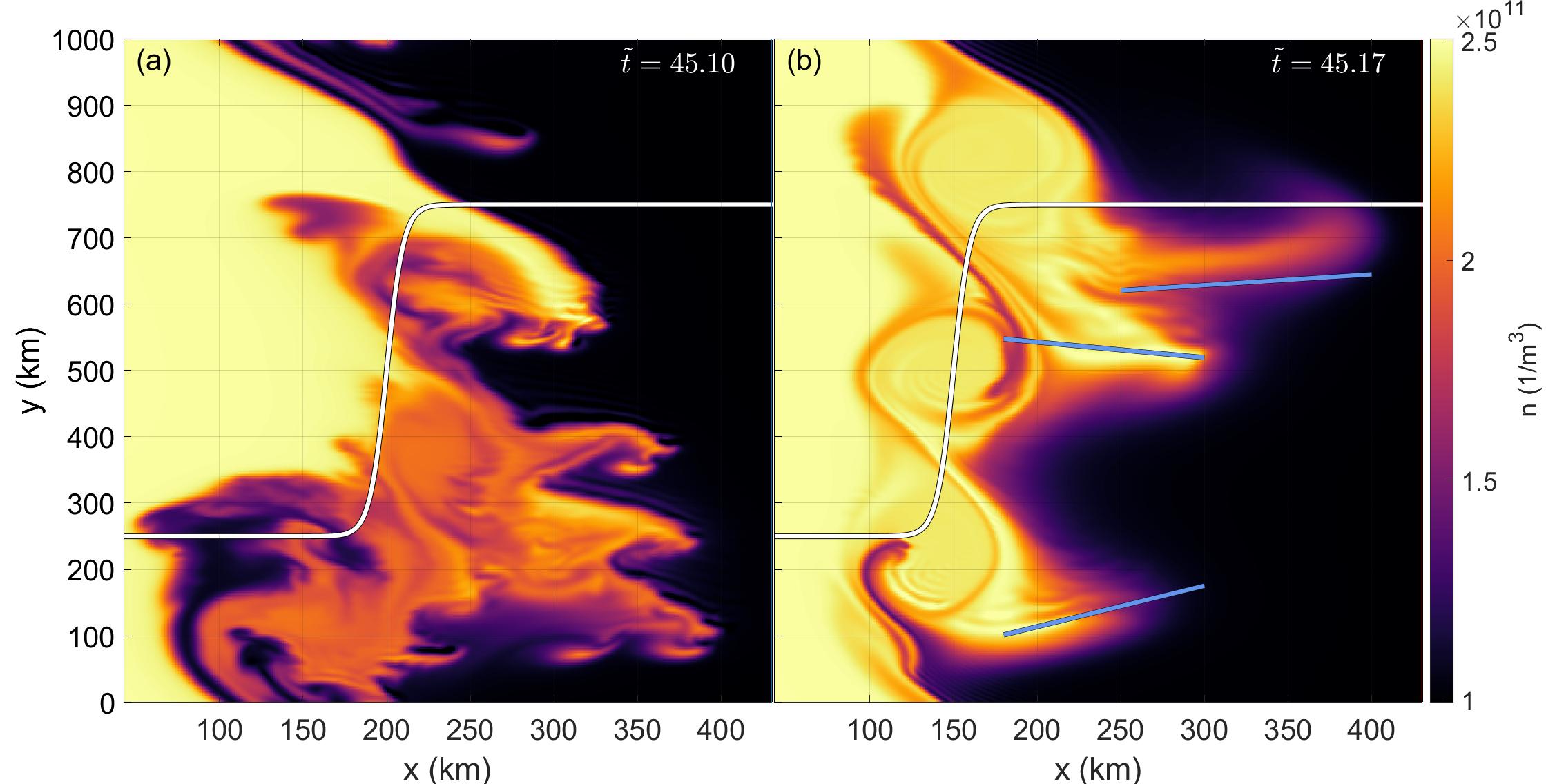}
	\caption{Density color plots showing KHI development using the same parameters
		as Figure~\ref{f:shear} but with an $\mathbf{\hat{x}}$ direction
		neutral wind of 500 m/s. The neutral wind has an enhancing effect
		on the KHI growth in both cases. The KHI is 
		the dominant instability in Panel (a). 
		Panel (b) shows the GDI growing on the right side of the KHI. Light blue lines show the 
		expected growth directions for different density gradient directions for the secondary GDI.
			\label{f:shearUx}
	}
\end{figure}

Figure~\ref{f:shearUx} shows the late stage instability development with the neutral wind. 
The time is presented non-dimensionally as $\tilde{t} = |t V_0 / L_N^g |$.
The neutral wind has an enhancing effect on the KHI which is seen by the significantly
larger amplitude for both cases as compared to Figure~\ref{f:shear} despite 
being at similar times. However, the neutral wind does not impact the transition
of the KHI to the GDI with Figure~\ref{f:shearUx}(a), similar to Figure~\ref{f:shear}(a),
being dominated by the KHI. Figure~\ref{f:shearUx}(b) shows much clearer ``finger-like" structures
suggesting the growth of the GDI. 
Similar to Figure~\ref{f:shear}(b), 
the density gradient direction can vary from $\pi/2$ to $\pi$ due to
the generally circular nature of the KHI vortex which
leads to different optimal GDI growth directions.
Three blue lines
indicate the different GDI growh directions. Starting from the top line,
the density gradient directions used are $\theta_g=3\pi/4$, $\pi/2$, and $\pi$.
Note that in this case, because of the addition of the neutral wind, $\theta_e=-0.4636$.

\begin{figure}[!htb]  
	\includegraphics[width=\linewidth]{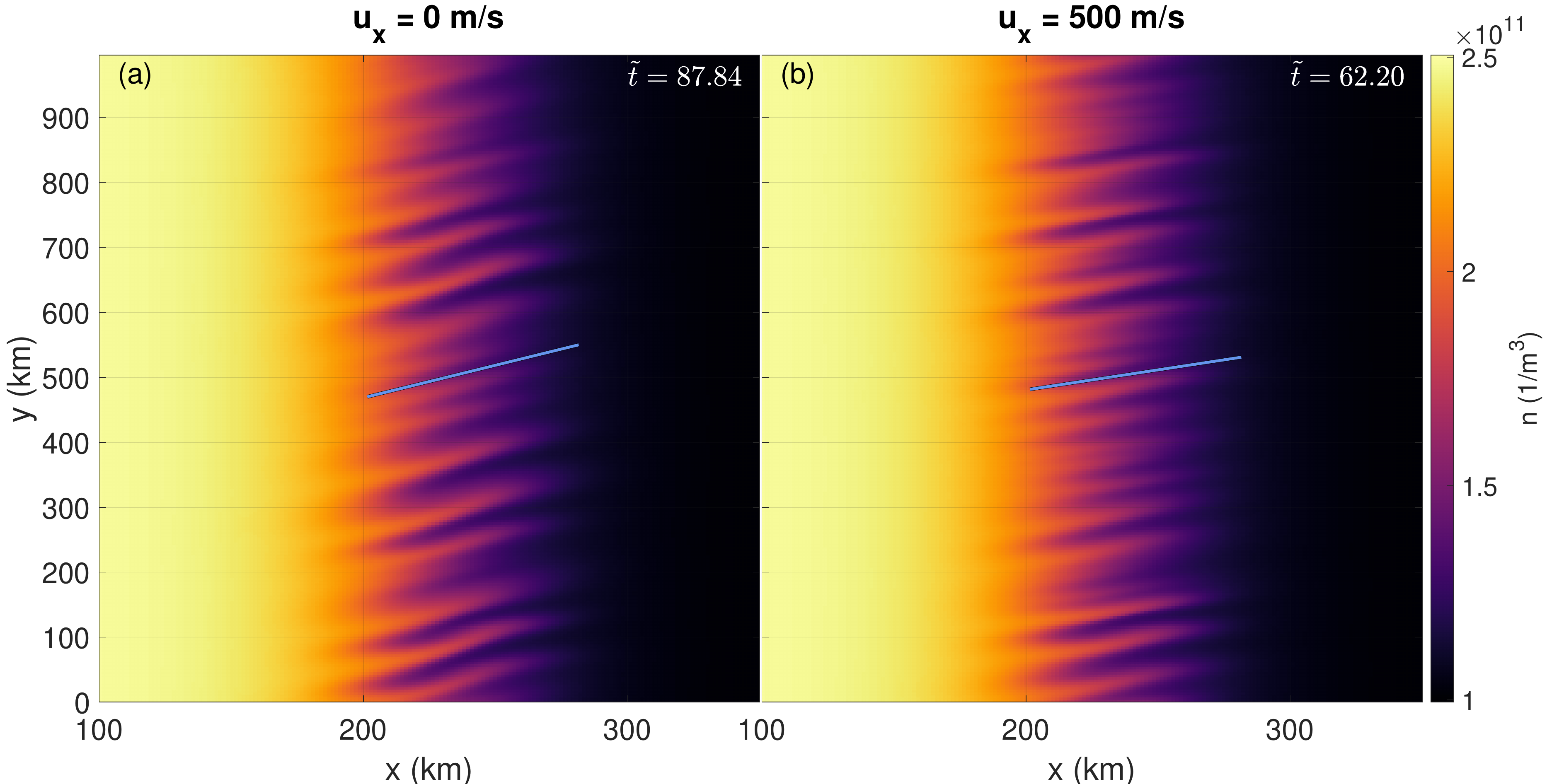}
	\caption{Panels (a) and (b) show the density color plots using the equivalent
		homogeneous applied electric field to Figures~\ref{f:shear}(b) and
		\ref{f:shearUx}(b), respectively. The blue line indicates the direction of optimal GDI growth.
		The growth of the GDI takes significantly longer than the growth of the GDI
		shown in Figures~\ref{f:shear}(b) and \ref{f:shearUx}(b).
			\label{f:neutralWindEquiv}
	}
\end{figure}

To show that the GDI is secondary to the KHI in the cases of 
Figures~\ref{f:shear}(b) and \ref{f:shearUx}(b), a set of simulations
are run using those same parameters 
but with $V_0=0$ m/s, $u_y=1000$ m/s, and a resolution of $1024 \times 256$ grid points. 
This neutral wind drives the equivalent electric field
that causes the background $\mathbf{E}\times\mathbf{B}$ drift but is constant
throughout the entire domain. Figure~\ref{f:neutralWindEquiv} shows results
with the time normalized as $\tilde{t} = |t u_y / L_N^g|$, which in this case is equivalent to
the time normalizations in Figures~\ref{f:shear}(b) and \ref{f:shearUx}(b).
The GDI grows in this low collisionality regime but at a much slower rate
compared to Figures~\ref{f:shear}(b) and \ref{f:shearUx}(b).
Therefore, the GDI in Figures~\ref{f:shear}(b) and \ref{f:shearUx}(b) is a secondary
instability that is seeded by the primary KHI.

\section{Discussion} \label{s:discussion}

Previous works have described formulae for the 
optimal GDI growth rate directions\cite{keskinen1982nonlinear,makarevich2014symmetry}.
Eq.~\ref{eq:thetak} provides a generalized form as a function of $R$,  $\theta_g$,
and $\theta_e$. 
Ref.~\onlinecite{keskinen1988nonlinear} examines this problem from a qualitative
perspective noting that GDI growth is tilted in the
generally expected direction for an angled electric field. 
Figure~\ref{f:GDIkDir}
provides a more comprehensive and detailed analysis showing
that Eq.~\ref{eq:thetak} accurately predicts the optimal direction of
GDI growth. This provides a useful diagnostic tool for both simulation and observation. 
In simulations, the direction of any GDI growth can be predicted prior to running the simulation.
For observation, the direction of observed GDI growth can provide information about possible unknowns such 
as the electric field direction, density gradient direction, or collisionality.

Several parameters have been examined to understand in which cases the GDI
or KHI might grow. Figure~\ref{f:shearUx} shows that 
the neutral wind has no impact on the transition between the GDI and the KHI.
Figures~\ref{f:shear} and~\ref{f:shearUx} show that for certain velocity profiles
at low $R$, the GDI develops as a secondary instability, based on Figure~\ref{f:neutralWindEquiv}, in the nonlinear phase
of the KHI. Linear theory and growth rates for this type of transition and secondary growth are
highly nontrivial. Furthermore, the dominating instability in these cases is still the KHI.
Figure~\ref{f:GDIKHIV1e4} shows that
the background velocity magnitude plays only a minor role
despite the velocity being the 
approximate upper limit of what occurs in the ionosphere.
Therefore, the collisional parameter $R$ is the main factor
in determining which instability dominates, as shown in Figures~\ref{f:GDIKHI} and~\ref{f:GDIKHIV1e4}.
From a physical perspective, changing $R$ can also be thought of 
as changing the altitude because the neutral density is a strong function of 
altitude.
Data from the IRI \cite{bilitza2018iri}, NRLMSISE-00 \cite{picone2002nrlmsise},
and IGRF \cite{thebault2015international} empirical ionospheric models are used
to calculate $R$ as a function altitude. The input parameters used are for the
date of May 2, 2013 at 20:00 local time at a geographic latitude and longitude
of $57.5^\circ$ and $40^\circ$ respectively. Since the composition of 
the ionosphere and the thermosphere are also functions of altitude, a weighted average
is used based on each species' density to obtain effective masses and collision cross-sections.
The ion and neutral particle collision cross-sections are obtained assuming hard sphere collisions
using the appropriate species' Van der Waals radius\cite{bondi1964van,batsanov2001van}.
The electron radius is assumed to be much smaller than the ion and neutral particle radii and is therefore set to 0 m.

\begin{figure}[!htb]  
	\includegraphics[width=\linewidth]{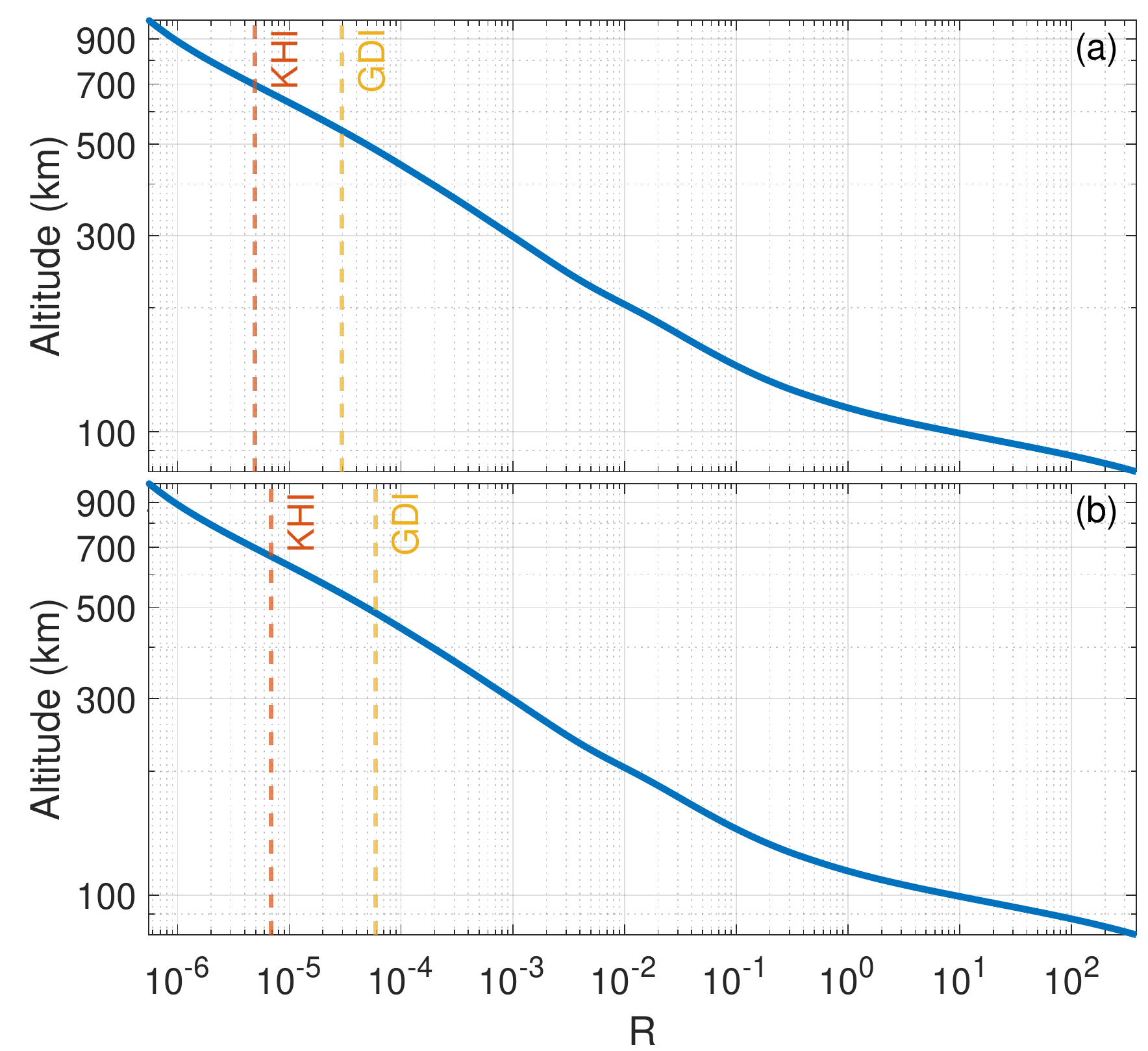}
	\caption{An example calculation of $R$ as a function of altitude
		using data from the IRI \cite{bilitza2018iri}, NRLMSISE-00 \cite{picone2002nrlmsise},
		and IGRF \cite{thebault2015international} models. As altitude increases,
		$R$ decreases. The red and orange dashed lines represent the $R$ the altitude limits for the KHI
		and GDI respectively. Panel (a) sets the limits based on the results from Figure~\ref{f:GDIKHI}
		and Panel (b) sets the limits based on the results from Figure~\ref{f:GDIKHIV1e4}.
		Above the red dashed line, the KHI will dominate; below the
		orange dashed line, the GDI will dominate. In the middle, both instabilities co-exist in some combination.
			\label{f:RvAlt}		
	}
\end{figure}

Figure~\ref{f:RvAlt} shows the results of the 
example calculation of $R$ (blue line). Because of the effect of the
neutral density, $R$ is found to decrease with increased altitude. 
Vertical red and orange dashed lines are plotted separating different regions of 
instablity behavior based on the background density
and velocity profiles considered. 
For altitudes above the red dashed line, the KHI dominates, i.e.,
would grow first; for altitudes below the orange dashed line,
the GDI dominates. In the intermediate region, both instabilities grow in tandem.
Note that the blue curve is the same for Figures~\ref{f:RvAlt}(a-b); the 
purpose of the two plots is to show
how the altitude limits of the transition from the GDI to KHI change with differing
background velocity shear.
Figure~\ref{f:RvAlt}(a) determines the different regions of dominance based on the results
shown in Figures~\ref{f:GDIKHI}(a) and \ref{f:GDIKHI}(c). 
Figure~\ref{f:RvAlt}(b) determines the different regions of dominance based on the results
shown in Figures~\ref{f:GDIKHIV1e4}(a) and \ref{f:GDIKHIV1e4}(c). 
Figure~\ref{f:RvAlt}(b) shows that increased
velocity shear causes the transition region to occur at a lower altitude.
The effect of decreasing plasma density with altitude is not expected to impact these results
as both the GDI\cite{makarevich2014symmetry} and the KHI\cite{chandrasekhar1961hydrodynamic,keskinen1988nonlinear}
are impacted only by the density gradient scale length or density ratio and not the magnitude.

Note that Figure~\ref{f:RvAlt} is just an example calculation using simulation parameters
from Figures~\ref{f:GDIKHI} and~\ref{f:GDIKHIV1e4}. Additionally, the
empirical parameters are found for a specific geographic location and time.
Thus, Figure~\ref{f:RvAlt} does not apply to every case in the ionosphere.
The key point is that there exists a region in altitude where the dominant instability
will transition from the GDI to the KHI in certain sheared $\mathbf{E}\times\mathbf{B}$ 
flows. Subauroral polarization streams (SAPS) \cite{foster2002saps} have been observed to exhibit
density irregularities\cite{ledvina2002first,mishin2003electromagnetic,foster2004millstone,oksavik2006first,mishin2008irregularities}.
Two instruments that have observed these are the Super Dual Auroral 
Radar Network (SuperDARN)\cite{oksavik2006first} and DMSP satellites \cite{mishin2003electromagnetic,mishin2008irregularities}.
SuperDARN radars operate at altitudes of approximately 300 km\cite{chisham2007decade} whereas
DMSP satellites have orbits of about 800 km\cite{mishin2008irregularities}. Based on
Figure~\ref{f:RvAlt} and with the correct SAPS velocity profiles,
the DMSP satellites would observe the KHI as the dominant instability
while the SuperDARN radars would observe the GDI as the dominant instability. 
Figure~\ref{f:RvAlt}(b) shows that even for the upper limit of ionospheric velocity
magnitudes, the DMSP satellites and SuperDARN radars would still
only capture the KHI and GDI, respectively.

\section{Summary and Conclusions} \label{s:conclusions}

This paper has studied the gradient drift (GDI) and Kelvin-Helmholtz (KHI)
instabilities using a model that solves a perturbed set of governing equations
to understand parameter regimes in which each instability might dominate
in a sheared $\mathbf{E}\times\mathbf{B}$ flow.

An analysis of the GDI linear theory provides a 
prediction on the fastest growing wavenumber direction. 
Simulations that vary the geometric conditions of the electric field and 
density gradient demonstrate the predictive capability providing a useful
diagnostic tool for simulation and observation.

For the density and velocity profiles considered in this work,
when modifying the collisional parameter, $R$, it is found that the KHI
is the dominating instability at $R$ lower than \num{4.62e-6} and the GDI is the dominating
instability at $R$ greater than \num{2.95e-5}. 
For intermediate $R$ values, both instabilities exist in tandem.
It is also found that the KHI is more sensitive to changes in
the background velocity. For higher velocity cases, it is found that the KHI
dominates for $R$ lower than \num{6.89e-6} and the GDI dominates
for $R$ larger than \num{5.90e-5}. This suggests that for higher velocities,
the entire transition region occurs at a lower $R$ or, from a physical
perspective, a lower altitude.

Velocity shear has a stabilizing effect on the GDI\cite{perkins1975velocity,huba1983shearF,huba1983shearE}.
For a background velocity profile sufficiently far from the density gradient,
a secondary GDI is formed that extends out of the KHI vortices. The inclusion
of an applied electric field serves to increase the KHI amplitude as opposed to causing the GDI
to dominate.

Since the collisionality is the primary factor
in determining whether the KHI or the GDI are
dominant, an example
calculation of $R$ as a function of altitude is provided. Data are obtained from 
the IRI \cite{bilitza2018iri}, NRLMSISE-00 \cite{picone2002nrlmsise},
and IGRF \cite{thebault2015international} models. 
It is found that above a certain altitude, the KHI is the dominating instablity. Below another altitude,
the GDI is the dominating instability. The implications are that with certain background density and velocity
profiles that exist over large altitude ranges, the dominating instabilities observed by satellite versus ground based systems about the 
same phenomenon (in this case subauroral polarization streams) may be significantly different.

\begin{acknowledgments}
	This work was supported by NASA under grant number NASAMAG16\_2-0050 and the Bradley Department of Electrical Engineering at Virginia Tech.
\end{acknowledgments}

\section*{Data Availability Statement}
The data that support the findings of this study are available from the corresponding author upon reasonable request.

\appendix

\section{Model Derivation \label{a:model_deriv}} 

This section presents the derivation of the
model, which is adapted from a theoretical framework 
presented in Ref.~\onlinecite{keskinen2004midlatitude}.
In addition, inertial terms are included to better study
the KHI. The primary novelty is the development of 
the perturbed model to be able to study turbulence
development in a wide array of physically relevant background
profiles.

The continuity, momentum, and energy equations
(Eqs.~\ref{eq:cont}-\ref{eq:temp}), rewritten here
for convenience, are
\begin{gather}
	\frac{ \partial n_\alpha }{ \partial t } + 
	\nabla \cdot ( n_\alpha \mathbf{V}_\alpha ) 
	= D \nabla^2 n_\alpha \label{eq:contApp} \\
	n_\alpha \Big( \frac{\partial}{\partial t} + \mathbf{V}_\alpha 
	\cdot \nabla \Big) \mathbf{V}_\alpha = 
	\frac{q_\alpha n_\alpha}{m_\alpha} ( 
	\mathbf{E} + \mathbf{V}_\alpha \times \mathbf{B}) \nonumber \\
	- \frac{\nabla P_\alpha}{m_\alpha} - 
	n_\alpha \nu_{\alpha n} ( \mathbf{V}_\alpha
	- \mathbf{u}) \label{eq:mtmApp} \\
	\frac{3}{2} n_\alpha k_B 
	\frac{\partial T_\alpha}{\partial t} 
	+ \frac{3}{2} n_\alpha k_B 
	\mathbf{V}_\alpha \cdot \nabla T_\alpha 
	+ n_\alpha k_B T_\alpha 
	\nabla \cdot \mathbf{V}_\alpha = 0. \label{eq:tempApp}
\end{gather}
The momentum equations, Eq.~\ref{eq:mtmApp},
can be solved for the velocity, $\mathbf{V}_\alpha$, in ordered terms
of $\nu_{\alpha n}/\Omega_{c \alpha}$ and $(\partial/\partial t
+ \mathbf{V} \cdot \nabla) / \Omega_{c \alpha}$. In the \textit{F} 
region ionosphere, these terms are considered small and thus only the zeroth
and first order velocities are retained.

The model considers 2D motion perpendicular to the magnetic field. 
Taking the cross product of the electron momentum equation, Eq.~\ref{eq:mtmApp} with $q_e=-e$,
and the magnetic field and solving for $\mathbf{V}_e$ yields
\begin{multline}
	\mathbf{V}_e = \mathbf{V}_e^0 + \mathbf{V}_e^1 + H.O.T. = 
	\underbrace{\frac{\mathbf{E} \times \mathbf{B} }{B^2} 
	+ \frac{\nabla P_e \times \mathbf{B} }{e n_e B^2}}_{\mathbf{V}_e^0}   \\
	+ \frac{1}{B} \frac{\nu_{en}}{ \Omega_{ce}}  \Big( \mathbf{V}_e - \mathbf{u} \Big) \times \mathbf{B}
	+ \frac{1}{B}\frac{1}{ \Omega_{ce}} \Big( \frac{\partial }{\partial t} +
	\mathbf{V}_e \cdot \nabla \Big) \mathbf{V}_e \times \mathbf{B},
	\label{eq:veMid1}
\end{multline}
where $H.O.T.$ stands for higher order terms. 
The zeroth order electron velocity, denoted as $\mathbf{V}_e^0$ by
the underbrace in Eq.~\ref{eq:veMid1}, is a summation
of the $\mathbf{E}\times\mathbf{B}$ and diamagnetic drifts. The remaining terms are of first
order or higher.

The zeroth order ion velocity is needed to calculate the first
order electron velocity. 
Similar to how Eq.~\ref{eq:veMid1} is obtained, the ion
velocity is
\begin{multline}
	\mathbf{V}= \mathbf{V}_i^0 + \mathbf{V}_i^1  + H.O.T. = 
	\underbrace{\frac{\mathbf{E}\times\mathbf{B}}{B^2}
	- \frac{\nabla P_i \times \mathbf{B}}{ e n_i B^2}}_{\mathbf{V}_i^0}  \\
	- \frac{1}{B}\frac{\nu_{in}}{\Omega_{ci}}  \Big( \mathbf{V}_i - \mathbf{u} \Big)  \times \mathbf{B}
	- \frac{1}{B}\frac{1}{ \Omega_{ci}} \Big( \frac{\partial}{\partial t} 
	+ \mathbf{V}_i \cdot \nabla \Big) \mathbf{V}_i \times \mathbf{B}.
	\label{eq:viMid1}
\end{multline}

The electron and ion velocities that appear on the right hand sides of
Eqs.~\ref{eq:veMid1} and \ref{eq:viMid1} can also be
decomposed into zeroth, first, and higher order terms, which
are all multiplied by terms of the form
$\nu_{\alpha n}/\Omega_{c \alpha}$ or $(\partial/\partial t
+ \mathbf{V} \cdot \nabla) / \Omega_{c \alpha})$. Therefore, only
the zeroth order terms remain as the first order terms become higher order
terms. Thus, the first order electron and ion velocities are
\begin{widetext}
\begin{equation}
	\mathbf{V}_e^1  =
	\frac{\nu_{en}}{B\Omega_{ce}} \Big(  \frac{\mathbf{E} \times \mathbf{B} }{B^2} 
	+ \frac{\nabla P_e \times \mathbf{B} }{e n_e B^2}  - \mathbf{u} \Big) \times \mathbf{B}  
	+ \frac{1}{B\Omega_{ce}} \Big( \frac{\partial }{\partial t} +
	\Big[  \frac{\mathbf{E} \times \mathbf{B} }{B^2} 
	+ \cancel{\frac{\nabla P_e \times \mathbf{B} }{e n_e B^2}}  \Big] \cdot \nabla \Big) \Big(  \frac{\mathbf{E} \times \mathbf{B} }{B^2} 
	+ \cancel{\frac{\nabla P_e \times \mathbf{B} }{e n_e B^2}}  \Big) \times \mathbf{B}
\end{equation}
\begin{equation}
	\mathbf{V}_i^1 =
- \frac{\nu_{in}}{B \Omega_{ci}}  \Big(  \frac{\mathbf{E} \times \mathbf{B} }{B^2} 
- \frac{\nabla P_i \times \mathbf{B} }{e n_i B^2}   - \mathbf{u} \Big)  \times \mathbf{B}
- \frac{1}{B \Omega_{ci}} \Big( \frac{\partial}{\partial t} 
+ \big[ \frac{\mathbf{E} \times \mathbf{B} }{B^2} 
- \cancel{\frac{\nabla P_i \times \mathbf{B} }{e n_i B^2} }  \big]\cdot \nabla \Big) 
\Big( \frac{\mathbf{E} \times \mathbf{B} }{B^2} 
- \cancel{\frac{\nabla P_i \times \mathbf{B} }{e n_i B^2} } \Big)\times \mathbf{B}.
\end{equation}
\end{widetext}
For the \textit{F} region phenomena considered in this work,
the diamagnetic drift term is much smaller than the $\mathbf{E}\times\mathbf{B}$ drift terms. Thus,
the diamagnetic terms are not considered in the convective derivative. After simplification,
the zeroth order velocities are 
\begin{align}		
	\mathbf{V}_e^0 &= \frac{\mathbf{E}\times\mathbf{B}}{B^2} 
	+ \frac{ \nabla P_e \times \mathbf{B} }{ e B^2 n_e}
	\label{eq:ve0App}\\
	\mathbf{V}_i^0 &= \frac{\mathbf{E}\times\mathbf{B}}{B^2} 
	- \frac{ \nabla P_i \times \mathbf{B} }{ e B^2 n_i},
	\label{eq:vi0App}
\end{align}
and the first order velocities are
\begin{align}		
	\mathbf{V}_e^1 &= -\frac{\nu_{en}}{\Omega_{ce}} 
	\frac{\nabla P_e}{e n_e B} -
	\frac{\nu_{en}}{B \Omega_{ce}} 
	\Big( \mathbf{E} + \mathbf{u} \times \mathbf{B} \Big) \nonumber \\
	&\ - \frac{1}{B\Omega_{ce}} 
	\Big( \frac{\partial}{\partial t} + 
	\mathbf{V}_{\mathbf{E}\times\mathbf{B}} \cdot \nabla \Big)
	\mathbf{E} \label{eq:ve1App}\\
	\mathbf{V}_i^1 &=  -\frac{\nu_{in}}{\Omega_{ci}} 
	\frac{\nabla P_i}{e n_i B} +
	\frac{\nu_{in}}{B \Omega_{ci}} 
	\Big( \mathbf{E} + \mathbf{u} \times \mathbf{B} \Big) \nonumber \\
	&\ + \frac{1}{B\Omega_{ci}} 
	\Big( \frac{\partial}{\partial t} + 
	\mathbf{V}_{\mathbf{E}\times\mathbf{B}} \cdot \nabla \Big)
	\mathbf{E}. \label{eq:vi1App}
\end{align}
Note that the inertia is primarily driven by the ions.

The zeroth order velocities, Eqs.~\ref{eq:ve0App} and \ref{eq:vi0App},
are used in the continuity and energy equations. The diamagnetic
drift term does not play a role in the continuity equation due
to the spatially constant magnetic field; thus, only the $\mathbf{E}\times\mathbf{B}$
drift advects the plasma. Furthermore, because the $\mathbf{E}\times\mathbf{B}$ drift is 
incompressible, the continuity equation can be written as
\begin{equation}
	\frac{ \partial n }{ \partial t } 
	- \frac{\nabla \phi \times \mathbf{B}}{B^2}
	\cdot \nabla n
	= D \nabla^2 n, \label{eq:contFinalApp} 
\end{equation}
where the electric field is $\mathbf{E} = - \nabla \phi$ where $\phi$ is 
the electric potential.

Assuming quasineutrality ($n_i=n_e=n$), the current closure
equation is 
\begin{equation}
	\nabla \cdot \mathbf{J} = \nabla \cdot \Big[ n e \big(\mathbf{V}_i - \mathbf{V}_e \big) \Big] = 0 ,
	\label{eq:JApp}
\end{equation}
where $\mathbf{J}$ is the current density. Subsituting Eqs.~\ref{eq:ve0App} to \ref{eq:vi1App}
into Eq.~\ref{eq:JApp} and solving for the temporal derivative term yields
\begin{multline}
	\nabla \cdot \Big( n \nabla 
	\frac{\partial \phi}{\partial t}  \Big) 
	=  \Big( \frac{1}{\Omega_{ci}} 
	+ \frac{1}{\Omega_{ce}} \Big)^{-1} 
	\bigg[ - \frac{\nu_{in}}{e\Omega_{ci}} \nabla^2 P_i 
	+ \frac{\nu_{en}}{e\Omega_{ce}} \nabla^2 P_e \\
	+ \Big( \frac{\nu_{in}}{\Omega_{ci}}
	+ \frac{\nu_{en}}{\Omega_{ce}}\Big) 
	\Big( \mathbf{u}\times\mathbf{B} \cdot \nabla n  
	-\nabla \cdot [ n \nabla \phi] \Big)\bigg] \\
	- \nabla \cdot \Big( n \mathbf{V}_{\mathbf{E}\times\mathbf{B}} \cdot \nabla \nabla \phi\Big).
	\label{eq:divJ0App}
\end{multline}

For the problems considered in this paper, the energy equations play
a negligible role in the instability development. While they are still 
solved, the emphasis is placed on
Eqs.~\ref{eq:contFinalApp} and \ref{eq:divJ0App}.

For problems in which the background dynamics are significantly slower than 
turbulence development, a perturbed model can be developed that
maintains a background that is constant in time and only evolves perturbations.
Each of the variables, $\phi$, $n$, $T_i$, and $T_e$, are split into
background and perturbed quantities, e.g., $\phi = \phi_{bg}+ \phi_p$.
These terms are substituted into 
Eqs.~\ref{eq:contFinalApp} and \ref{eq:divJ0App}.
The equations are then expanded using the distributive
property. Any terms that contain only background quantities are removed,
and thus, only the perturbed terms remain.

For the continuity equation,
Eq.~\ref{eq:contFinalApp}, only the temporal derivative and the 
numerical diffusion source term need to be split in this way. The residual
term automatically satisfies this condition because all of the background profiles
used in Section~\ref{s:results} are only functions of $x$ and the magnetic
field is in the $\mathbf{\hat{z}}$ direction. Thus, the term 
$( \nabla \phi_{bg}\times \mathbf{B} ) \cdot \nabla n_{bg} $ is implicitly set to 0. 
As a simple example, the diffusion term is split into background and perturbed components as
\begin{equation}
	D \nabla^2 n = D \Big( \cancel{\nabla^2 n_{bg}}
	- \nabla^2 n_p \Big) = D \nabla^2 n_p .
	\label{eq:diffApp}
\end{equation}
Because the diffusion term is linear, the final perturbed result looks the same 
as the original term, except that the perturbed density is used instead of the total density.
Similarly, the temporal derivative in the continuity equation is linear, resulting
in a perturbed continuity equation of 
\begin{equation}  
	\frac{\partial n_p}{\partial t} - 
	\frac{\nabla \phi \times \mathbf{B}}{B^2}
	\cdot \nabla n
	= D \nabla^2 n_p
	\label{eq:contPertApp}
\end{equation}

The current closure equation, Eq.~\ref{eq:divJ0App},
is highly nonlinear, making it more complicated to obtain
its perturbed equivalent. Since the background electric potential is assumed to be
constant in time, $\partial \phi_{bg} / \partial t$ is 0.
The pressure terms are linear and decomposed similar to the
diffusion term, as shown in Eq.~\ref{eq:diffApp}. Similarly,
the neutral wind term is also linear and easily decomposed. 
However, the terms $\nabla \cdot ( n \nabla \phi)$ and
$\nabla \cdot (n \mathbf{V}_{\mathbf{E}\times\mathbf{B}} \cdot \nabla \nabla \phi )$
are nonlinear, with the latter being highly so. The decomposition of the first term
is 
\begin{align}
	\nabla \cdot \big( n \nabla \phi \big)_p &=
	\nabla \cdot \Big[ \big( n_{bg} + n_p \big) 
	\nabla \big( \phi_{bg} + \phi_p \big) \Big] \nonumber \\
	&= \nabla \cdot \big( \cancel{ n_{bg} \nabla \phi_{bg}} + 
	n_{bg} \nabla \phi_p + n_p \nabla \phi_{bg} + n_p \nabla \phi_p \big) \nonumber \\
	&= \nabla \cdot \big( n_p \nabla \phi + n_{bg} \nabla \phi_p \big).
\end{align}
Note how the term $n_{bg} \nabla \phi_{bg}$ consists
of only background terms and is thus, not considered in the perturbed 
model. 
The same method is used to obtain the perturbed form
of the remaining highly nonlinear term. 
The full perturbed current closure equation is
\begin{widetext}
	\begin{multline} 
		\nabla \cdot \Big( n \nabla 
		\frac{\partial \phi_p}{\partial t} \Big) =
		\Big( \frac{1}{\Omega_{ci}} 
		+ \frac{1}{\Omega_{ce}} \Big)^{-1} 
		\bigg[ - \frac{\nu_{in}}{e\Omega_{ci}} \nabla^2 P_{i_p}  
		+ \frac{\nu_{en}}{e\Omega_{ce}} \nabla^2 P_{e_p} 
		+ \Big( \frac{\nu_{in}}{\Omega_{ci}}
		+ \frac{\nu_{en}}{\Omega_{ce}}\Big) 
		\Big( \mathbf{u}\times\mathbf{B} \cdot \nabla n_p   
		-\nabla \cdot [ n_p \nabla \phi + n_{bg} \nabla \phi_p] \Big)\bigg] \\
		- \nabla \cdot \Big( n_{bg} \mathbf{V}_{\mathbf{E}\times\mathbf{B}_{bg}} \cdot \nabla \nabla \phi_p + n_{bg} \mathbf{V}_{\mathbf{E}\times	
			\mathbf{B}_p}\cdot \nabla \nabla \phi + n_p \mathbf{V}_{\mathbf{E}\times\mathbf{B}} \cdot \nabla \nabla \phi \Big).
		\label{eq:divJpertApp}
	\end{multline}
\end{widetext}

\section{Comparison to Linear Theory \label{a:lin}}

Single mode slab geometry simulations of
the GDI and KHI are run using
the model derived in Appendix~\ref{a:model_deriv}.
The simulation results are compared to the linear theory
of these two instabilities in order to benchmark the model.
For all of these simulations, a single mode in the $\mathbf{\hat{y}}$ direction
is initialized. The modes are modified through each run by changing the 
domain length in $y$.

For both instabilities, the domain integrated electric field energy ($E^2 = E_x^2 + E_y^2$)
is calculated as a function of time. A series of best fit exponentials of
the form, $a \exp ( 2 \gamma t) $ are found to obtain the linear 
regime, where $a$ is some initial value, $t$ is the time, and
$\gamma$ is the desired growth rate. This growth rate is calculated for multiple simulations
with different excited wavenumbers to obtain the growth rate as a function of the wavenumber. 
These simulation growth rates are then compared to those from linear theory.

The GDI simulations are run 
using $512 \times 16$ grid points
with a domain length of 
$L = \SI{750}{km} \times 2\pi/k_y$ where $k_y$ is varied
from \SI{6.28e-3}{km^{-1}} to \SI{1.61}{km^{-1}}.
The background density profile is defined
by Eq.~\ref{eq:n} with
$n_0=10^{11}$ \si{m^{-3}},
$a_1 = -0.75$,
$a_2 = 0.75$,
$c = 1$,
$x_{N_1}=220$ km,
$x_{N_2}=525$ km,
and $L_N=75$ km.
The background velocity profile is set to 0. 
The neutral wind is 
$\mathbf{u} = -500 \mathbf{\hat{x}}$~m/s.
The background magnetic field is $B=\SI{5e-5}{T}$. 
The neutral and ion species are atomic and ionic oxygen respectively,
with a neutral number density of $n_n=10^{14}$~\si{m^{-3}}, which
corresponds to $R=\num{9.84e-5}$. The numerical diffusion constant
is $D=10^3$ \si{m^2/s}.

\begin{figure}[!htb]  
	\includegraphics[width=\linewidth]{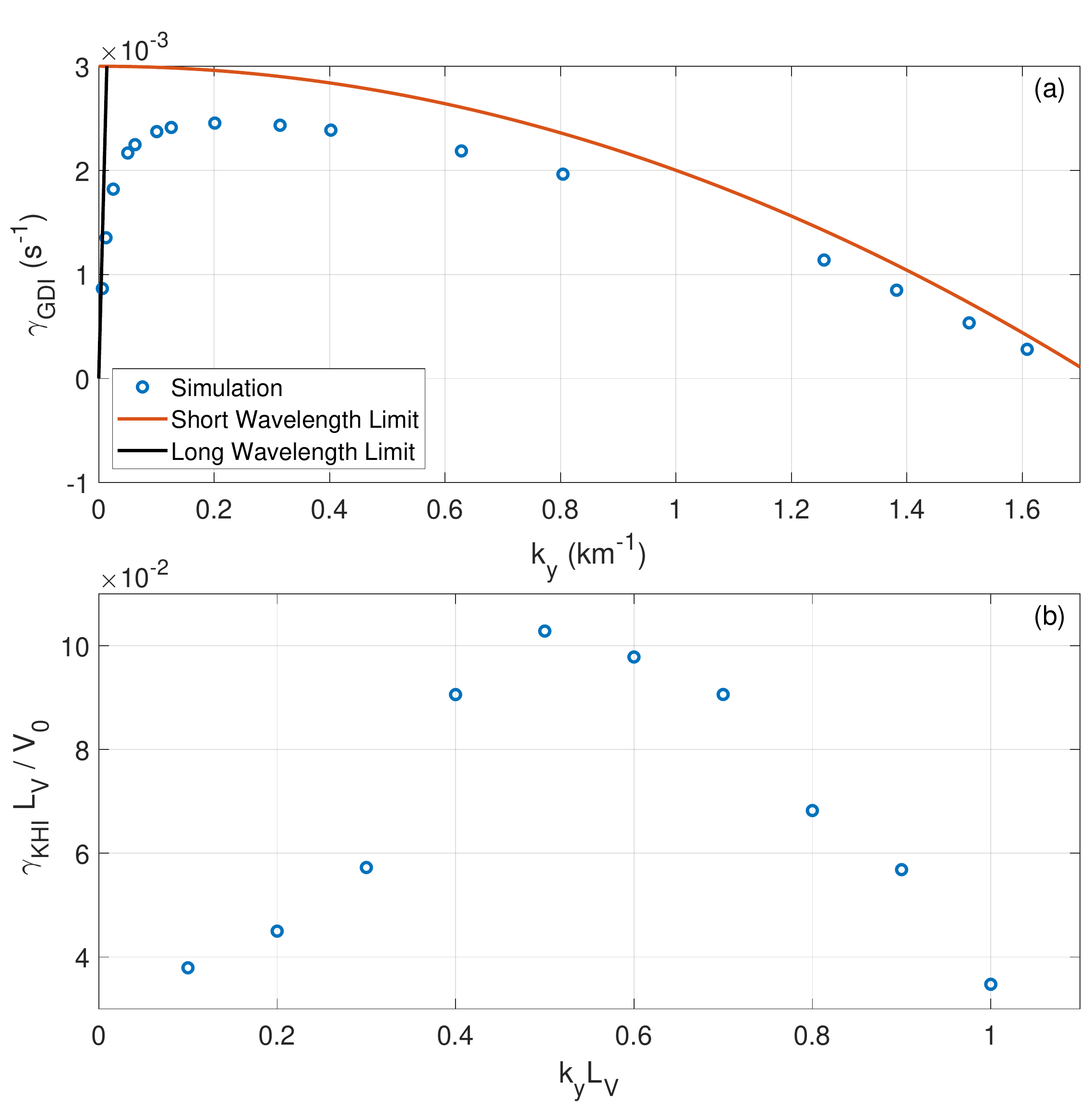}
	\caption{Plots of the growth rate as a function of wavenumber for 
		single mode GDI (a) and KHI (b) simulations.
		Panel (a) shows the GDI growth rate, where the circles 
		are the simulation growth rate, the red line is the short
		wavelength limit (Eq.~\ref{eq:GDIshortApp}), and the black
		line is the long wavelength limit (Eq.~\ref{eq:GDIlongApp}).
		The results indicate that the simulation growth rate tends to 
		approach the appropriate limit as $k_y$ approaches 0 or $\infty$.
		Panel (b) shows the KHI growth rate calculated from the simulations.
		This growth rate is compared to Figure 3 from Ref.~\onlinecite{keskinen1988nonlinear},
		showing reasonable agreement of the location of peak growth and magnitude of
		the growth rate.
		\label{f:singleMode}		
	}
\end{figure}

Figure~\ref{f:singleMode}(a) shows the GDI growth rate as a function
of the wavenumber. The blue circles are the growth rates calculated from 
the simulation.
The red line shows the short wavelength limit of the GDI growth rate, which,
in this geometry and accounting for the numerical diffusion, is\cite{makarevich2014symmetry}
\begin{equation}
	\gamma_{GDI} = - \frac{u_x}{L_N^g} - D k_y^2.
	\label{eq:GDIshortApp}
\end{equation}
The black line shows the long wavelength limit of the GDI growth rate,
which is\cite{huba1983long}
\begin{equation}
	\gamma_{GDI} = - A k u_x,
	\label{eq:GDIlongApp}
\end{equation}
where $A=(n_1 - n_2)/(n_1+n_2)$ is the Atwood number
with $n_1$ being the density of the heavier fluid
and $n_2$ being the density of the lighter fluid.
Figure~\ref{f:singleMode} shows
that the simulation growth rate tends to
Eq.~\ref{eq:GDIshortApp} as $k\rightarrow \infty$ 
and Eq.~\ref{eq:GDIlongApp} as $k\rightarrow 0$.
Therefore, there is good agreement between the simulation 
and the linear theory.

The KHI simulations are run using $512 \times 16$ 
grid points. Many of the KHI length parameters are based
on the length $L_V$ from Eq.~\ref{eq:vTanh}, which is used
to obtain the background velocity profile, with
$L_V=10$ km, $V_0 = 1000$ m/s,
$x_{V_1}=4 L_V$, and
$x_{V_2} = 8 L_V$. The domain size
is set to $20 L_V \times 2 \pi / k_y$, where
$k_y$ is defined such that the value $k_y L_V$ is varied
from 0.1 to 1.0. Eq.~\ref{eq:n} is used for
the density profile with 
$n_0=10^{11}$ \si{m^{-3}},
$a_1 = -1$,
$a_2 = 1$,
$c = 1$,
$x_{N_1} = 4 L_V$, 
$x_{N_2} = 16 L_V$, and
$L_N=L_V$.
The background neutral wind is set to 0.
The background magnetic field is $B=\SI{5e-5}{T}$. 
The neutral and ion species are atomic and ionic oxygen respectively,
with a neutral number density of $n_n=10^{11}$~\si{m^{-3}}, which
corresponds to $R=\num{9.87e-8}$. 
The numerical diffusion constant
is $D=0$ \si{m^2/s}.

Figure~\ref{f:singleMode}(b) shows the simulation growth rate, normalized
by $V_0/L_V$, for the KHI
as a function of the wavenumber, normalized by $L_V$.
The normalized ion-neutral collision frequency is
$\nu_{in} L_V / V_0 = \num{2.96e-4}$. 
This result can be compared to Figure 3 from Ref.~\onlinecite{keskinen1988nonlinear}. 
Note that due to the complicated nonlocal analysis, the growth rate does not
have an analytical form\cite{keskinen1988nonlinear}.
Figure~\ref{f:singleMode}(b) shows that
the location of peak growth in
occurs at about $k_y L_V = 0.5$
which is consistent with Figure 3 from Ref.~\onlinecite{keskinen1988nonlinear}.
The magnitude of the growth rate is on the same order of magnitude
as the results from Ref.~\onlinecite{keskinen1988nonlinear}. 
Thus, there is reasonable agreement between the simulation results and
linear theory.

Thus, the GDI and KHI reasonably match linear theory
in conditions that are already well understood. This paper
shows that there are regimes which have not been studied in the literature.
Future work constitutes conducting a nonlocal linear analysis to investigate,
from a theoretical perspective, the GDI and the KHI in combination.

\bibliography{reference}

\end{document}